\documentclass[12pt,letterpaper]{article}
\usepackage[usenames, dvipsnames]{xcolor}
\usepackage{jcapmod}

\usepackage{tocloft}
\usepackage[]{todonotes}
\usepackage{verbatim}
\usepackage[utf8]{inputenc}
\usepackage{mathrsfs}
\usepackage[shortlabels]{enumitem}

\usepackage{pgf}
\usepackage{tikz-cd}
\usetikzlibrary{shapes,arrows}
\usetikzlibrary{calc}
\usepackage{pgfplots}
\pgfplotsset{compat=1.12}
\usepackage{tikz-3dplot}

\usepackage{tikz}
\usepackage{setspace,caption}
\usepackage{lipsum}
\usepackage[below]{placeins}
\usetikzlibrary{matrix,arrows,calc}
\makeatletter
\newsavebox\myboxA
\newsavebox\myboxB
\newlength\mylenA

\definecolor{cornellRed}{HTML}{B31B1B}
\definecolor{cornellBlue}{HTML}{0068AC}
\definecolor{cornellGreen}{HTML}{6EB43F}

\usepackage{bbm} 					
\usepackage{slashed} 				
\usepackage{graphicx}				
\usepackage{subcaption}			
\usepackage{psfrag}				
\usepackage{tensor}				
\usepackage{fouridx}				
\usepackage{bm}					
\usepackage{mdframed}				
\usepackage{multirow}				
\usepackage{soul}					
\usepackage{bbold}				
\usepackage{multicol}				
\usepackage{tikz-cd}
\usepackage{rotating}

\usetikzlibrary{arrows}

\tikzset{
commutative diagrams/.cd,
arrow style=tikz,
diagrams={>=latex}}

\usepackage{amsmath}
\usepackage{amssymb}
\usepackage{amsfonts}
\usepackage{mathtools}

\usepackage{feynmf}
\usepackage{marvosym}

\usepackage{import}

\newcommand*\xoverline[2][0.75]{%
    \sbox{\myboxA}{$\m@th#2$}%
    \setbox\myboxB\null
    \ht\myboxB=\ht\myboxA%
    \dp\myboxB=\dp\myboxA%
    \wd\myboxB=#1\wd\myboxA
    \sbox\myboxB{$\m@th\overline{\copy\myboxB}$}
    \setlength\mylenA{\the\wd\myboxA}
    \addtolength\mylenA{-\the\wd\myboxB}%
    \ifdim\wd\myboxB<\wd\myboxA%
       \rlap{\hskip 0.5\mylenA\usebox\myboxB}{\usebox\myboxA}%
    \else
        \hskip -0.5\mylenA\rlap{\usebox\myboxA}{\hskip 0.5\mylenA\usebox\myboxB}%
    \fi}
\makeatother

\newcommand{\cF}{\mathcal{F}}

\newcommand{\cK}{\mathcal{K}}
\newcommand{\cL}{\mathcal{L}}
\newcommand{\cM}{\mathcal{M}}

\newcommand{\cQ}{\mathcal{Q}}
\newcommand{\cR}{\mathcal{R}}

\newcommand{\cV}{\mathcal{V}}

\newcommand{\eV}{\mathrm{eV}}

\definecolor{cobalt}{RGB}{44, 98, 120}
\definecolor{celadon}{rgb}{0.67, 0.88, 0.69}
\definecolor{dm}{cmyk}{.20, 0, .30, 0}
\definecolor{burgundy}{rgb}{0.5, 0.0, 0.13}
\definecolor{plotBlue}{RGB}{94, 130, 181}

\hypersetup{
  colorlinks,
  citecolor=Violet,
  linkcolor=cobalt,
  urlcolor=Blue}

\DeclareSymbolFontAlphabet{\mathbb}{AMSb}

\newif\iffastcompile

\fastcompiletrue

\iffastcompile
\newcommand{\cl}[1]{}
\newcommand{\lm}[1]{}
\newcommand{\md}[1]{}
\newcommand{\ab}[1]{}
\else
\newcommand{\cl}[1]{\todo[color=burgundy!30, size=\scriptsize, bordercolor=burgundy!30]{CL: #1}}
\newcommand{\lm}[1]{\todo[color=dm!90, size=\scriptsize, bordercolor=dm!90]{LM: #1}}
\newcommand{\md}[1]{\todo[color=blue!30, size=\scriptsize, bordercolor=blue!50]{MD: #1}}
\newcommand{\ab}[1]{\todo[color=blue!30, size=\scriptsize, bordercolor=blue!50]{AB: #1}}

\fi

\newcommand{\email}[1]{\href{mailto:#1}{#1}}

\ProvideTextCommandDefault{\Dbar}{%
\leavevmode\lower.5ex\rlap{\hskip-.07em\accent"16}D%
}

\begin{document}
	\newcommand{\main}{.}
\begin{titlepage}

\setcounter{page}{1} \baselineskip=15.5pt \thispagestyle{empty}
\setcounter{tocdepth}{2}

\bigskip\

\vspace{1cm}
\begin{center}
{\fontsize{22}{28} \bfseries The Kreuzer-Skarke Axiverse}

\end{center}

\vspace{0.45cm}

\begin{center}
\scalebox{0.95}[0.95]{{\fontsize{14}{30}\selectfont Mehmet Demirtas,$^{a}$ Cody Long,$^{b}$ Liam McAllister,$^{a}$ and Mike Stillman$^{c}$ \vspace{0.25cm}}}

\end{center}

\begin{center}

\textsl{$^{a}$Department of Physics, Cornell University, Ithaca, NY 14853, USA}\\
\textsl{$^{b}$Department of Physics, Northeastern University, Boston, MA 02115, USA}\\
\textsl{$^{c}$Department of Mathematics, Cornell University, Ithaca, NY 14853, USA}\\

\vspace{0.25cm}

\email{\tt md775@cornell.edu, co.long@northeastern.edu,\\ mcallister@cornell.edu, mike@math.cornell.edu }
\end{center}

\vspace{0.6cm}
\noindent
We study the topological properties of Calabi-Yau threefold hypersurfaces at large $h^{1,1}$.
We obtain two million threefolds $X$ by triangulating polytopes from the Kreuzer-Skarke list, including all polytopes with $240 \le h^{1,1}\le 491$.  We show that the K\"ahler cone of $X$ is very narrow at large $h^{1,1}$, and as a consequence, control of the $\alpha^{\prime}$ expansion in string compactifications on $X$ is correlated with the presence of ultralight axions.  If
every effective curve has volume $\ge 1$ in string units, then the typical volumes of irreducible effective
curves and divisors, and of $X$ itself, scale as $(h^{1,1})^p$, with $3\lesssim p \lesssim 7$ depending on the type of cycle in question.
Instantons from branes wrapping these cycles are thus highly suppressed.

\noindent
\vspace{2.1cm}

\noindent\today

\end{titlepage}
\tableofcontents\newpage


\section{Introduction}

As a step toward understanding general properties of quantum gravity in four spacetime dimensions, one can study compactifications of weakly-coupled string theories on six-manifolds whose curvatures are small in string units.  Understanding what is possible in such compactifications can shed light on what is possible in quantum gravity.

Bounds on topological or geometric properties of a class of compactifications can
imply interesting statements about the corresponding effective theories, such as bounds on the number of fields, the rank of the gauge group, or the diameter of moduli space.  Moreover, an understanding of generic properties of compactifications can inform low-energy model building, e.g.~the fact that typical Calabi-Yau threefolds have scores or hundreds of moduli and axions suggests considering inflationary and dark sectors that might be discarded as excessively complicated from a bottom-up perspective.

The Kreuzer-Skarke database of four-dimensional reflexive polytopes \cite{Kreuzer:2000xy} is a fount of data on Calabi-Yau compactifications.  A fine,  regular, star triangulation (FRST) of any of the 473,800,776 polytopes $\Delta^{\circ}$ in the list determines a toric variety $V$, in which a generic anticanonical hypersurface is a smooth Calabi-Yau threefold $X$.  However, only the most elementary data, such as the Hodge numbers of $X$, can be obtained directly from the database without computation.  To more fully characterize a compactification on $X$, one needs to compute and manipulate an FRST of $\Delta^{\circ}$.

A key measure of the difficulty of this computation is the number $N$ of relevant lattice points in $\Delta^{\circ}$, or equivalently the
Picard number $h^{1,1}$ of $X$: when $X$ is favorable, $N = h^{1,1}+4$.
The number of possible triangulations of a given $\Delta^{\circ}$, and the complexity of each triangulation, both grow rapidly with $N$.  Most publicly-available
software, such as {\tt{Sage}}, is effective only for $N \lesssim 10$.
Through a major computational effort, Altman et al.~obtained the data of all compactifications with $N \le 10$ \cite{Altman}. However, there have been few studies, none systematic, of compactifications on Calabi-Yau threefold hypersurfaces with $h^{1,1} \gg 1$. This is a critical gap in our understanding: the Kreuzer-Skarke list contains threefolds with $h^{1,1}$ as large as 491, and vast numbers of triangulations --- corresponding to potentially-distinct threefolds --- are possible at $h^{1,1}\gg 1$.
To the best of our knowledge, most Calabi-Yau threefold hypersurfaces have not yet been examined.

In this work we initiate a study of Calabi-Yau threefold hypersurfaces with large Picard number.  We obtain FRSTs of 2,031,335 reflexive polytopes with $2 \le h^{1,1} \le 491$, including one triangulation for each polytope with $240\leq h^{1,1} \leq 491$.\footnote{Huang and Taylor have shown that all Hodge number pairs with $h^{1,1} \ge 240$ in the Kreuzer-Skarke list can be realized by elliptically fibered Calabi-Yau threefolds \cite{Wati}.}
We compute the Mori cones of the associated toric varieties $V$, and for $2 \le h^{1,1} \le 100$ we compute the intersection numbers of Calabi-Yau hypersurfaces $X \subset V$.

We first use these data to bound the K\"ahler cone ${\cal K}_X$ of $X$.
We know of no efficient algorithm to compute ${\cal K}_X$ directly in a hypersurface with $h^{1,1} \gtrsim 10$, so we instead place upper and lower bounds by computing cones containing ${\cal K}_X$, and contained in ${\cal K}_X$.  The K\"ahler cone ${\cal K}_V$ of $V$ obeys ${\cal K}_V \subset {\cal K}_X$, while a cone ${\cal K}_{\cap}$ associated to the intersections of divisors
$\widehat{D} \subset V$ (see \S\ref{sec:background} for a precise definition) obeys ${\cal K}_X \subset {\cal K}_{\cap}$.

Equipped with bounds on the K\"ahler cone of $X$, we examine the $\alpha^{\prime}$ expansion in a compactification of string theory\footnote{For specificity one can imagine type IIB string theory on an orientifold of $X$, but most of what follows is purely geometric, and applies, \emph{mutatis mutandis}, in other string theories.} on $X$.  For the $\alpha^{\prime}$ expansion to be under control, we will require
that every holomorphic curve $\Sigma$
on $X$ has volume obeying
\begin{equation} \label{equ:zero}
\mathrm{Vol}(\Sigma) \ge (2\pi)^2 c\,\alpha^{\prime} \equiv c\, \ell_s^2\,,
\end{equation}
with $c$ a dimensionless constant, so that worldsheet instantons wrapping $\Sigma$ give corrections to the effective action
$\lesssim e^{-2\pi c}$.
Although we will suppose that $c$ is of order unity, one can immediately extend our findings to any desired numerical value of $c$.  We henceforth set $\ell_s^2 = 1$ and $c = 1$, and so the constraint \eqref{equ:zero} reads
\begin{equation} \label{equ:first}
\mathrm{Vol}(\Sigma) \ge 1\,.
\end{equation}
We argue in \S\ref{sec:background} that (\ref{equ:first}) is a useful proxy for control of perturbative and nonperturbative corrections in the $\alpha^{\prime}$ expansion.

The requirement that (\ref{equ:first}) holds for all $\Sigma$ typically implies that \emph{some}
irreducible
holomorphic curves have volumes $\gg 1$.  Moreover, some
irreducible effective divisors have even larger volumes, and the total threefold volume $\cal{V}$ is larger still.  At first glance these trends appear unsurprising:
the number of nonvanishing triple intersection numbers $\kappa_{ijk}$ must grow with $h^{1,1}$, and so too should $\mathcal{V}=\frac{1}{6}\kappa_{ijk}t^i t^j t^k$, where $t_{i},\, i=1,\ldots,h^{1,1}$, are the volumes of a basis of $H_2(X,\mathbb{Z})$.
However, obtaining the intersection numbers $\kappa_{ijk}$ for a hypersurface with $h^{1,1} \gg 1$ is computationally expensive, so prior studies of this point have been very limited.
In this work we precisely quantify the growth of curve, divisor, and threefold volumes with $h^{1,1}$: our computation of ${\cal K}_{\cap}$ leads to lower bounds on these quantities.
The volumes grow far more quickly with $h^{1,1}$ than can be accounted for by the growth of the intersection numbers alone.
We find that
there are only $\mathcal{O}(h^{1,1})$ nonvanishing intersection numbers in each geometry, with mean size independent of $h^{1,1}$, yet we find that $\cal{V}$ grows as $(h^{1,1})^p$ with $p \sim 7$: see \S\ref{sec:results}.

The primary cause of this rapid growth of volumes is the \emph{narrowness of the K\"ahler cone}.
The K\"ahler cone conditions enforce inequalities relating the various curve volumes, and with increasing $h^{1,1}$, this effect becomes more pronounced, because the number of inequalities grows.  Intuitively, the K\"ahler cone becomes very narrow for $h^{1,1} \gg 1$, so to be well-separated from every wall one must be very far from the origin of the cone.

One physical consequence of this finding is that requiring control of the $\alpha^{\prime}$ expansion, in the sense of (\ref{equ:first}), typically leads to ultralight axions, unless $h^{1,1}$ is small.  As an example, in a compactification of type IIB string theory on an orientifold of a hypersurface $X$, the Ramond-Ramond four-form $C_4$ gives rise to axion fields that are massless to all orders in perturbation theory, and acquire mass from Euclidean D3-branes.  Suitable holomorphic four-cycles (i.e., suitable effective divisors) support superpotential contributions \cite{Witten:1996bn}, which are well-understood, while non-holomorphic four-cycles can support contributions to the K\"ahler potential.  We find that for typical geometries in our ensemble,
every basis constructed from generators of the cone of effective divisors (cf.~\S\ref{sec:masses}) contains elements with volume $\gtrsim (h^{1,1})^3$: see Figure \ref{fig:divvol}.  Thus, superpotential couplings\footnote{We argue in Appendix \ref{sec:nonholomorphic} that contributions to the axion masses from K\"ahler potential instantons are plausibly comparably suppressed.} give extremely small masses to some of the axions.  In every geometry in our ensemble with $h^{1,1} > 22$, the lightest axion is essentially massless, with $m < 10^{-33}\,\rm{eV}$.

An important caveat is that our finding of rapid growth of volumes with $h^{1,1}$ is a consequence of
the requirement (\ref{equ:first}).  It is possible that $\alpha^{\prime}$ corrections to the four-dimensional action are small in some cases even if some effective curves have volumes violating (\ref{equ:first}).  Constraining this possibility would be worthwhile, but would likely require advances in computing perturbative corrections in the $\alpha^{\prime}$ expansion.
Moreover, our qualitative results would be unaffected unless most curves can be made small in string units.

We also study the radius
of the axion fundamental domain for each geometry in our ensemble.
Understanding whether super-Planckian displacements of an inflaton field can occur in well-controlled compactifications is a pressing problem, and one way forward is to search for geometries in which the axion field space has radius $\mathcal{R} \gg M_{\rm{pl}}$.
Prior work in \cite{Long:2016jvd} has shown that $\mathcal{R} \lesssim \mathcal{O}(1)$ in every Calabi-Yau hypersurface with $h^{1,1} \le 4$.  Here we extend the analysis of \cite{Long:2016jvd} to $2 \times 10^6$ hypersurfaces with $5 \le h^{1,1} \le 100$.  We show that $\mathcal{R} \lesssim M_{\rm{pl}}$ for most of the geometries in our ensemble.
However, in a small fraction of cases we cannot exclude the possibility of radii $\mathcal{R} \gg M_{\rm{pl}}$ in the parameter regime where (\ref{equ:first}) holds and the $\alpha^{\prime}$ expansion is well-controlled.  Obtaining definitive results in these intriguing cases would require advances in computing the K\"ahler cones of Calabi-Yau hypersurfaces \emph{per se}, rather than just the K\"ahler cones of the corresponding ambient toric varieties.

The organization of this note is as follows.  In \S\ref{sec:background} we review basic facts about the K\"ahler cone of a Calabi-Yau hypersurface in a toric variety.
In \S\ref{sec:str} we introduce the notion of a stretched K\"ahler cone, and in \S\ref{sec:masses} we explain how upper bounds on axion masses can be obtained by
computing cycle volumes in an appropriate stretched K\"ahler cone.
In \S\ref{sec:methodology} we describe our algorithm for computing the K\"ahler cones, and approximations to the K\"ahler cones, in an ensemble of Calabi-Yau threefold hypersurfaces constructed from the Kreuzer-Skarke database.  We present our results in \S\ref{sec:results}.
In \S\ref{sec:implications} we explore the implications of our findings for the axion mass spectrum in type IIB compactifications.
We conclude in \S\ref{sec:conclusions}.  Although our findings directly involve the volumes of holomorphic cycles, in Appendix \S\ref{sec:nonholomorphic} we discuss how instantons wrapping non-holomorphic volume-minimizing chains could be governed to good approximation by the growth of volume that we establish in the holomorphic case.

\section{The Effective, K\"ahler, and Mori Cones} \label{sec:background}

In this section we recall the definitions and basic properties of the effective cone, the K\"ahler cone, and the Mori cone of a projective algebraic variety $X$, and we explain how to compute approximations to these cones when $X$ is a Calabi-Yau threefold hypersurface in a toric variety.   From the data of these convex cones one can read off properties of the effective theory arising
in a string compactification on $X$.

\subsection{The effective cone} \label{effsec}

Let $X$ be a projective algebraic variety of complex dimension $n$.  A Weil divisor $D$ on $X$ is a finite formal sum of irreducible codimension-one subvarieties $D_A$,
\begin{equation}
D = \sum_A n_A D_A \qquad n_A \in \mathbb{Z}\,.
\end{equation}
The divisor $D$ is called \emph{effective} if the $n_A$ are all nonnegative.
We define the \emph{effective cone} $\mathrm{Eff}(X)$ to be the convex cone in $H_{2n-2}(X,\mathbb{R})$ spanned by the classes of effective divisors.

The relevance of the effective cone is that a Euclidean D3-brane wrapping a divisor $D$ in an orientifold of a Calabi-Yau threefold $X$ can contribute to the superpotential only if $D$ is effective.  Intuitively, effective divisors consist of finite collections of irreducible holomorphic hypersurfaces, each of which can support BPS D-branes.

\subsubsection{Effective divisors of a Calabi-Yau hypersurface} \label{CYeffsec}

Let $\Delta^\circ$ be a four-dimensional reflexive polytope.
An FRST of $\Delta^\circ$ defines a fan that corresponds to a simplicial toric fourfold $V$.  The generic anticanonical hypersurface $X \subset V$ is a smooth Calabi-Yau threefold \cite{Batyrev}.

Each lattice point $v^I$ on the boundary of $\Delta^\circ$ corresponds to a homogeneous toric coordinate $x^I$, whose vanishing defines a \emph{prime toric divisor} $\widehat{D}_I$.
The prime toric divisors are irreducible effective divisors on $V$.
A subset $\{v^A\} \subset \{v^I\}$ of the points on the boundary of $\Delta^\circ$
are not interior to 3-faces (facets) of $\Delta^{\circ}$, but instead lie in faces of dimension $\le 2$.  Each such lattice point $v^A$ not interior to a facet corresponds to a prime toric divisor that intersects $X$ transversely.
The restriction to $X$ then defines a divisor $D_A\subset X$,
\begin{equation}
D_A := \widehat{D}_A \cap X\,,
\end{equation} that is effective on $X$.
Points interior to facets, on the other hand, define divisors of $V$ that do not intersect a generic Calabi-Yau hypersurface $X$.  In triangulating $\Delta^\circ$ we may therefore ignore lattice points interior to facets; such a triangulation corresponds to a maximal projective crepant partial (MPCP) desingularization, in the sense of ~\cite{cox1999mirror}. We will restrict ourselves to such partial desingularizations.

In general, $D_A$ may be a reducible divisor on $X$, even though $\widehat{D}_A$ is irreducible on $V$.  This occurs if and only if $v^A$ corresponds to a point in the strict interior of a 2-face $f \subset \Delta^\circ$, and $\ell^{*} (f^\circ)>0$, where $\ell^{*} (f^\circ)$ is the number of lattice points in the strict interior of the dual face $f^\circ \subset \Delta$. The condition that all of the prime toric divisors $\widehat{D}_A$ on $V$ that intersect $X$ in fact restrict to irreducible divisors on $X$ is thus
\begin{equation} \label{favcond}
\sum\limits_{f \subset \Delta^\circ} \ell^{*} (f) \ell^{*} (f^\circ) = 0 \, ,
\end{equation}
where the sum is over all 2-faces $f \subset \Delta^\circ$ . A polytope obeying \eqref{favcond} is called \emph{favorable}, and by extension we refer to the associated $V$ and $X$ as being favorable.

For simplicity we will confine our attention to the case where $X$ is favorable, though we expect the results of our analysis to extend into the non-favorable regime.
For $X$ favorable, there are exactly $h^{1,1}(X)+4$ prime toric divisors $\widehat{D}_A$.
We call
\begin{equation}
\{D_A\} := \{\widehat{D}_A \cap X \}\, \qquad A=1,\ldots,h^{1,1}(X)+4\,
\end{equation} the \emph{inherited prime toric divisors} on $X$.

The set $\{D_A\}$, $A=1,\ldots,h^{1,1}(X)+4$, provides a complete set of generators for $H_4(X, \mathbb{Z})$.  Since $\mathrm{dim} H_4(X, \mathbb{Q})=h^{1,1}(X)$, by reordering the $D_A$
we can ensure that $\{D_i\}$, $i=1,\ldots,h^{1,1}(X)$, is a basis for $H_4(X, \mathbb{Q})$.

\subsubsection{Inherited and autochthonous divisors} \label{inheritsec}

The inherited prime toric divisors $D_A$ of a Calabi-Yau threefold hypersurface $X \subset V$ are effective divisors on $X$ that are inherited from effective divisors on $V$.  We call the cone in $H_4(X,\mathbb{R})$ generated by the classes of the $\{D_A\}$ the \emph{inherited effective cone} $\mathrm{Eff}_{\iota}(X)$.  Clearly, $\mathrm{Eff}_{\iota}(X) \subseteq \mathrm{Eff}(X)$.  Because $V$ is specified by combinatorial data, it is straightforward to compute $\mathrm{Eff}_{\iota}(X)$.  However, in many cases $\mathrm{Eff}_{\iota}(X) \subsetneq \mathrm{Eff}(X)$: that is, there are effective divisors on $X$ that are not inherited from any effective divisor on $V$.  We call such a non-inherited divisor
an \emph{autochthonous} divisor.

In this work, we approximate $\mathrm{Eff}(X)$ by $\mathrm{Eff}_{\iota}(X)$. In particular, in computing axion masses in compactifications of type IIB string theory on $X$, we here consider only Euclidean D3-branes wrapping inherited effective divisors.
Autochthonous divisors of Calabi-Yau hypersurfaces are studied in our forthcoming work \cite{Picardwipmisc}.  Among other things, we show there that Euclidean D3-branes wrapping autochthonous divisors do not significantly affect the axion mass hierarchies found here, and so for present purposes it suffices to study the conceptually and computationally simpler inherited effective cone.

\subsection{The K\"ahler cone and the Mori cone}\label{kahlerdef}

Let $X$ be a projective algebraic variety, and let $J \in H^{1,1}(X,\mathbb{R})$ be a closed
(1,1)-form on $X$.  For a $k$-dimensional subvariety $U \subset X$, we define
\begin{equation}
\mathrm{Vol}_J(U):=\frac{1}{k!}\int_U \wedge^k J\,.
\end{equation}
We define the \emph{K\"ahler cone of $X$}, $\cK_X$, as the subset of $H^{1,1}(X,\mathbb{R})$ consisting of cohomology classes of K\"ahler forms $J$ on $X$, i.e.~$J$ such that $\mathrm{Vol}_J(U)>0$ for all subvarieties $U$.
The K\"ahler cone $\cK_X$, also called the \emph{ample} cone, is an open convex cone whose closure $\overline{\cK}_X$ is the cone of \emph{nef} (1,1) classes.\footnote{See \cite{demailly2004geometry} for a more detailed treatment.}

We next define the \emph{Mori cone} of $X$, $\mathcal{M}_X$, to be the cone in $H_2(X,\mathbb{R})$ generated by irreducible algebraic curves $C_a$ on $X$. (The Mori cone of $X$
is often denoted $\mathrm{NE}(X)$ in other parts of the literature.)
The K\"ahler cone and the Mori cone are related by
\begin{equation}\label{eq:dual}
\mathcal{M}^{\vee}=\overline{\cK}_X\,,
\end{equation} i.e.~the dual of the Mori cone is the closure of the K\"ahler cone.

When $X$ is a Calabi-Yau threefold hypersurface, the subvarieties of interest are the curves $C_a$, the divisors $D_A$, and the threefold itself. The volumes of these subvarieties are
\begin{equation}\label{eq:thevols}
	\begin{aligned}
	\mathfrak{t}_a :=& \mathrm{Vol}_J(C_a) = \int_{C_a} J\, , \\
	\tau_A:=& \mathrm{Vol}_J(D_A) = \frac{1}{2}\int_{D_A} J \wedge J\, , \\
	\cV:=& \mathrm{Vol}_J(X) = \frac{1}{6}\int_{X} J \wedge J \wedge J\, .
	\end{aligned}
\end{equation}
It is convenient to expand $J$ in terms of the Poincar\'e duals $[D_i]$ of the divisors $D_i$,
\begin{equation}
	J = t^i [D_i]\, .
\end{equation}
Defining
\begin{equation}  \label{intnumberdef}
	\begin{aligned}
	M_{ai} :=& \#C_a \cap D_i\, , \\
	\kappa_{Ajk} :=& \#D_A \cap D_j \cap D_k\, ,\\
	\kappa_{ijk} :=& \#D_i \cap D_j \cap D_k\, ,
	\end{aligned}
\end{equation}
the volumes \eqref{eq:thevols} are then written as
\begin{equation}
	\begin{aligned}
	\mathfrak{t}_a =& M_{ai} t^i\, ,\\
	\tau_A =& \frac{1}{2}\kappa_{Ajk} t^j t^k\, , \\
	\cV =& \frac{1}{6}\kappa_{ijk} t^i t^j t^k\, .
	\end{aligned}
\end{equation}
The $h^{1,1}$ K\"ahler parameters $t^i$, which are not necessarily positive when $J$ is inside the K\"ahler cone, should not be confused with the curve volumes $\mathfrak{t}_a$, which are positive for $J \in \cK_X$.

\section{The Stretched K\"ahler Cone} \label{sec:str}

One of the aims of this work is to determine the volumes of holomorphic submanifolds in $X$, when
every effective curve in $X$ is constrained to have volume $>1$, as in \eqref{equ:first}.
We therefore need to determine the cone of effective curves, i.e.~the Mori cone $\mathcal{M}_X$.

To our knowledge there is no established algorithm for computing $\mathcal{M}_X$, even for the well-studied ensemble of Calabi-Yau threefold hypersurfaces.
However,
we will identify two cones $\mathcal{M}_\mathrm{in}$ and $\mathcal{M}_\mathrm{out}$ that bound $\mathcal{M}_X$ on the inside and the outside, respectively, i.e.
\begin{equation}
\mathcal{M}_\mathrm{in} \subseteq \mathcal{M}_X \subseteq \mathcal{M}_\mathrm{out}\,,
\end{equation}
and it is these bounding cones that we will study.
The duals of these cones will then provide cones that bound $\cK_X$ on the outside and the inside, respectively: defining
$\overline{\mathcal{K}}_\mathrm{in}:=\mathcal{M}_\mathrm{out}^\vee$ and $\overline{\mathcal{K}}_\mathrm{out}:=\mathcal{M}_\mathrm{in}^\vee$,
and writing $\mathcal{K}_\mathrm{in}$ for the interior of $\overline{\mathcal{K}}_\mathrm{in}$, and
$\mathcal{K}_\mathrm{out}$
for the interior of $\overline{\mathcal{K}}_\mathrm{out}$, we have
\begin{equation}\label{inouteq}
\mathcal{K}_\mathrm{in} \subseteq \mathcal{K}_X \subseteq \mathcal{K}_\mathrm{out}\,.
\end{equation}
As we shall see, the K\"ahler cone $\mathcal{K}_V$ of the ambient toric variety $V$ can play the role of $\mathcal{K}_\mathrm{in}$, while a new cone, $\mathcal{K}_\cap$, provides the outer bound $\mathcal{K}_\mathrm{out}$~\cite{Cicoli:2018tcq}.

\subsection*{$\bm{\cK_V}$:}

Although computing $\mathcal{M}_X$ is challenging, the Mori cone $\mathcal{M}_V$ of the toric variety $V$ can be computed efficiently from the fan using an algorithm due to Berglund, Katz, and Klemm~\cite{Berglund:1995gd}, which is equivalent to the classical algorithm of Oda and Park~\cite{oda1991}.
By \eqref{eq:dual}, the dual of $\mathcal{M}_V$ is the closure $\overline{\cK}_V$ of the K\"ahler cone $\cK_V$ of $V$.
Restricting the K\"ahler parameters $t^i$ so that $t^i [D_i] \in \cK_V$
ensures that all holomorphic submanifolds of $V$ have positive volume, and therefore this restriction also guarantees that all holomorphic submanifolds of $X$ have positive volume. We therefore have
\begin{equation}
\cK_{V} \subseteq \cK_{X}\, .
\end{equation}

We remark that subvarieties of $V$ that correspond to simplices interior to facets do not intersect a generic $X$, and therefore any triangulations of $\Delta^\circ$ that differ only by simplices interior to facets define isomorphic Calabi-Yau hypersurfaces, but with different toric ambient spaces $V_\alpha$.
It is then natural to glue the K\"ahler cones  $\cK_{V_\alpha}$
together and define $\cK_\cup$~\cite{cox1999mirror}:
\begin{equation}
\cK_\cup := \underset{\alpha}{\bigcup} \, \cK_{V_\alpha}\, .
\end{equation}
However, such a process
appears prohibitively complicated at large $h^{1,1}$, and will not play a role in our analysis.

\subsection*{$\bm{\cK_\cap}$:}

Consider the following set of surfaces in $V$:
\begin{equation}
\{\widehat{S}_{AB}\} := \{\widehat{D}_A \cap \widehat{D}_B, A,B=1,\ldots,h^{1,1}+4, A\neq B\}\, .
\end{equation}
The intersection of any of the $\widehat{S}_{AB}$ with a generic anticanonical hypersurface $X$, if nonempty, is transverse and defines a corresponding curve in $X$,
\begin{equation}\label{cintform}
C_{AB} = D_A \cap D_B \subset X\qquad (A\neq B)\,.
\end{equation}
The curve $C_{AB}$ lies in $\mathcal{M}_X$, but in general not every element of $\mathcal{M}_X$ can be written in the form \eqref{cintform}.  Because the $\{C_{AB}\}$ are the curves inherited from intersections of distinct prime toric divisors, we call the $\{C_{AB}\}$ \emph{toric intersection curves}. The volumes of the toric intersection curves are
\begin{equation}
\mathrm{Vol}(C_{AB}) \equiv \mathfrak{t}_{AB}:= \int\limits_{D_A\cap D_B} J\, .
\end{equation}
We define the \emph{intersection cone} $\cK_\cap$ as the space of K\"ahler parameters $t^i$ for which the volumes $\cV$, $\tau_A$ and $\mathfrak{t}_{AB}$ are all positive:
\begin{equation}
\cK_\cap := \{J \,|\, \cV, \tau_A, \mathfrak{t}_{AB}>0\}\, .
\end{equation}
As these conditions are necessary, but in general not
sufficient,\footnote{In a few cases, $\cK_\cup =  \cK_\cap$ and we may therefore determine $\cK_{X}$ exactly, but this is far from generic.} to ensure that
$t^i [D_i] \in \cK_{X}$, we have the inclusions
\begin{equation}
\cK_V  \subseteq \cK_{X} \subseteq \cK_\cap\, .
\end{equation}

\subsection*{The stretched K\"ahler cone:}

In order to study the effect of demanding that all cycles satisfy the minimal volume constraint \eqref{equ:first}, we introduce the notion of a stretched K\"ahler cone.
Let $X$ be a projective algebraic variety, let $J \in H^{1,1}(X,\mathbb{R})$ be a closed
(1,1) form on $X$, and let $\mathcal{W}=\{W\}$ be a set of subvarieties $W\subset X$.
Given a number $c>0$, we define the $(c,\mathcal{W})$-\emph{stretched K\"ahler cone} of $X$,
\begin{equation}
\widetilde{\cK_{X}}[c,\mathcal{W}]:=\Bigl\{ J \in H^{1,1}(X,\mathbb{R}) \,\Bigl|\, \mathrm{Vol}_J(W) \ge c~~\forall~W \in \mathcal{W} \Bigr. \Bigr\}\,.
\end{equation}
The first stretched K\"ahler cone we consider is the \emph{stretched K\"ahler cone of X},
\begin{equation}
\widetilde{\cK_X}:=\widetilde{\cK_{X}}[1,\{C \in \mathcal{M}_X \}]\,.
\end{equation}
We next define the \emph{stretched intersection cone}
\begin{equation}\label{defkcap}
\widetilde{\cK_\cap}:=\widetilde{\cK_{X}}[1,\{C_{AB},D_A,X\}]\,,
\end{equation} as the region in which all intersection curves $C_{AB}$, all inherited prime toric divisors $D_A$, as well as $X$ itself, have volume $\ge 1$.
In all cases we have $\widetilde{\cK_X} \subseteq \widetilde{\cK_\cap}$, but because the curves $C_{AB}$ typically do not generate $\mathcal{M}_X$, we typically have $\widetilde{\cK_X} \subsetneq \widetilde{\cK_\cap}$.
Finally, noting that for favorable $X$, $H^{1,1}(V,\mathbb{R})$ can be naturally identified with $H^{1,1}(X,\mathbb{R})$,
we define the \emph{stretched K\"ahler cone of V},
\begin{equation}\label{eq:stretch2}
\widetilde{\cK_V}:=\widetilde{\cK_{V}}[1,\{\widehat{C} \in \mathcal{M}_V \}]\,,
\end{equation} i.e.~$\widetilde{\cK_V}$ is the subset of $H^{1,1}(X,\mathbb{R})\cong H^{1,1}(V,\mathbb{R})$ in which all curves $\widehat{C}$ on $V$ have volume $\ge 1$.\footnote{In a general computation of $\widetilde{\cK_V}$ using the algorithm of~\cite{Berglund:1995gd}, care would be needed to ensure that toric curves $\widehat{C}$ that can be singular in $V$ obey the constraint \eqref{equ:first} with $c=1$, rather than with some fractional $c$.  However, for our analysis it suffices to require that \emph{smooth} toric curves obey \eqref{equ:first}, and this is readily checked using \cite{Berglund:1995gd}.}

In a complete toric variety, any curve is rationally (and thus numerically) equivalent to an effective sum of toric curves~\cite{Reid1983}. A curve $\widehat{C} \subset X \subset V$ is also a curve in $V$,
and so in homology $\widehat{C}$
can be expressed as a non-negative integral linear combination of toric curves. It follows that $\widetilde{\cK_V} \subset \widetilde{\cK_X}$.

We have therefore bounded the stretched K\"ahler cone:
\begin{equation}\label{inout}
\widetilde{\cK_V} \subseteq \widetilde{\cK_X} \subseteq \widetilde{\cK_\cap}\, .
\end{equation}

\section{Axion Couplings}\label{sec:masses}

Consider a compactification of type IIB string theory on an orientifold\footnote{For simplicity we suppose here that $h^{1,1}_{-}=0$.} of a Calabi-Yau threefold hypersurface $X$.
The four-dimensional theory contains $h^{1,1}$ axions
from reduction of the Ramond-Ramond four-form $C_4$.
In this section we explain how the kinetic and potential couplings of the axion fields are computed from geometric data.

\subsection{Kinetic term}

In terms of a basis $\{D_i\}$, $i=1,\ldots,h^{1,1}$ for $H_4(X,\mathbb{Z})$, we define
\begin{equation}
\theta_i := \int_{D_i} C_4
\end{equation} to be the corresponding dimensionless axions.
The K\"ahler coordinates on K\"ahler moduli space are the complexified divisor volumes
\begin{equation}
T_i := \tau_i + i \theta_i\,,
\end{equation} with $\tau_i = \frac{1}{2}\int_{D_i} J\wedge J$, cf.~\eqref{eq:thevols}.  The axion kinetic term is then\footnote{Indices on $\tau_i$ and $\theta_i$ are raised with the identity matrix.}
\begin{equation}
\mathcal{L}_{\rm{kin}} = -\frac{M_{\rm{pl}}^2}{2} K_{ij}\partial^{\mu}\theta^i\partial_{\mu}\theta^j\,,
\end{equation} where the K\"ahler metric $K_{ij}$ is obtained from the K\"ahler potential $\mathscr{K}=-2\,\log\,\mathcal{V}$.

\subsection{Nonperturbative superpotential}

The axions are perturbatively massless and receive mass only  nonperturbatively, from instantons: specifically, from Euclidean D3-branes wrapping four-cycles.\footnote{Strong gauge dynamics on a stack of D7-branes wrapping a four-cycle can also produce a nonperturbative contribution to the axion potential.  Our considerations apply equally to Euclidean D3-branes and to D7-branes, but for simplicity of language we only refer to the former.}

The leading-order
bosonic action $S$ for a Euclidean D3-brane is given by the Dirac-Born-Infeld action plus an imaginary Chern-Simons term that provides the coupling to the axion (see e.g.~\cite{Blumenhagen:2009qh}).
Consider Euclidean D3-branes wrapping the four-cycles
\begin{equation}
\Sigma_\alpha := n_{\alpha}^{~i} D_{i}\,,
\end{equation} for some $n_{\alpha}^{~i} \in \mathbb{Z}$,~$\alpha=1,\ldots,\mathcal{N}$, and for some $\mathcal{N}>0$.
The action $S_\alpha$ of the Euclidean D3-brane wrapping $\Sigma_\alpha$ is then
\begin{equation}
S_\alpha = 2\pi\mathrm{Vol}(\Sigma_\alpha)+2\pi i\int_{\Sigma_\alpha}C_4 = 2\pi\mathrm{Vol}(\Sigma_\alpha)+2\pi i n_{\alpha}^{~i} \theta_i  \,.
\end{equation}
Although one can in principle consider Euclidean D3-branes wrapping any four-cycle $\Sigma_\alpha \in H_4(X,\mathbb{Z})$, the situation is best-understood when $\Sigma_\alpha$ is an effective divisor, i.e.~when
$[\Sigma_\alpha] \in \mathrm{Eff}(X)$: precisely in that case,
$\Sigma_\alpha$ is calibrated by the K\"ahler form $J$, and so obeys
\begin{equation}
\mathrm{Vol}(\Sigma_{\alpha}) = \frac{1}{2}\int_{\Sigma_{\alpha}} J \wedge J = n_{\alpha}^{~i} \tau_i\,,
\end{equation}
so that $S_{\alpha}=2\pi n_{\alpha}^{~i}(\tau_i+i\theta_i)=2\pi n_{\alpha}^{~i}T_i$.

If instead $[\Sigma_\alpha] \not\in \mathrm{Eff}(X)$, determining the volume of the minimum-volume representative of the class $[\Sigma_\alpha]$ is in general very difficult, as we explain in Appendix \ref{sec:nonholomorphic}.  Moreover, Euclidean D3-branes wrapping
a representative $\Sigma_\alpha$ of a class $[\Sigma_\alpha] \not\in \mathrm{Eff}(X)$ cannot contribute to the superpotential.  They may contribute to the K\"ahler potential, but such effects are not well understood.

For now we will focus on effective divisors, and we suppose that superpotential terms arise from Euclidean D3-branes wrapping the divisors
\begin{equation}\label{qwhichD}
D_{\alpha}:=q_{\alpha}^{~i} D_i  \in \mathrm{Eff}_{\iota}(X) \,,
\end{equation} for some $q_{\alpha}^{~i} \in \mathbb{Z}$,~$\alpha=1,\ldots,p$, and for some $p>0$.
The superpotential then takes the form \cite{Witten:1996bn,Kachru:2003aw}
\begin{equation} \label{kkltwis}
	W = W_0 + \sum_{\alpha} \mathcal{A}_{\alpha} \exp\big(-2\pi q_{\alpha}^{~i} T_i\big)
\end{equation} where $W_0$ is the classical flux superpotential \cite{Gukov:1999ya}.
The Pfaffians $\mathcal{A}_{\alpha}$ depend on the complex structure moduli, and will be set to unity in our analysis.
The axion potential can then be written as
\begin{equation}\label{axionvis}
\begin{aligned}
	V &= -\frac{8 \pi}{\cV^2} \Big[ \sum_{\alpha} q_{\alpha}^{~i} \tau_i W_0 e^{-2\pi q_{\alpha}^{~i} \tau_i} \cos{\big(2\pi q_{\alpha}^{~i} \theta_i\big)}\\
	 &+ \sum_{\alpha > \alpha '} \Big(\pi (K^{-1})_{ij} q_{\alpha}^{~i} q_{\alpha'}^{~j} +
	 (q_{\alpha}^{~i} + q_{\alpha'}^{~i}) \tau_i\Big) e^{-2\pi \tau_i (q_{\alpha}^{~i} + q_{\alpha'}^{~i} )} \cos{\big(2\pi\theta_i (q_{\alpha}^{~i} - q_{\alpha'}^{~i})\big)} \Big]
\end{aligned}
\end{equation}
We will make the conservative choice $W_0 \sim 1$: a smaller value of the
flux superpotential would make our upper bounds on axion masses more stringent.
Performing a $GL(h^{1,1},\mathbb{R})$ transformation $\phi_i= M_{\rm{pl}} M_{i}^{~j}\theta_j$ such that $\phi$ has canonical kinetic term, we arrive at
\begin{equation}
\mathcal{L} = -\frac{1}{2} \partial^{\mu}\phi_i\partial_{\mu}\phi^i-V(\phi)\,.
\end{equation}
The Hessian of the canonically-normalized axions is
\begin{equation}
\mathcal{H}_{ij}:= \frac{\partial^2}{\partial\phi_i\partial\phi_j}V(\phi)\,,
\end{equation} and we denote its eigenvalues by $h_1^2 \le \ldots \le h_{h^{1,1}}^2$.
The potential \eqref{axionvis} has a rich structure of minima and critical points, cf.~e.g.~\cite{Marsh:2011aa,Bachlechner:2012at,Aligned}, and finding the global
minimum numerically is expensive when $h^{1,1}\gg 1$ and $p \gg h^{1,1}$ (for $p$ slightly larger than $h^{1,1}$, which does not hold here, the methods of \cite{Aligned} could be used).  In the remainder, by \emph{axion masses-squared} we mean the Hessian eigenvalues $h_i^2$, evaluated at the origin $\vec{0}$,
i.e.~at $\theta_1=\theta_2=\cdots=\theta_{h^{1,1}}=0$.
By \emph{minimum axion mass-squared} we mean
\begin{equation}\label{mmindef}
m_{\rm{min}}^2 := \displaystyle \min_{i} |h^2_i(\vec{0})|\,.
\end{equation}
One should bear in mind that these quantities could change slightly upon shifting the axion vev to a minimum, but we have found no evidence for changes large enough to invalidate our parametric results.

\subsection{Geometric field ranges}
The effective Lagrangian for the axions is usefully rewritten as
\begin{equation}\label{eq:potential}
\cL = -\frac{M_{\rm{pl}}^2}{2} K_{ij}\partial^{\mu}\theta^i\partial_{\mu}\theta^j - \sum\limits_{a=1}^P \Lambda_a^4 \left( 1 - \text{cos}(Q_a^{~i} \theta_i) \right)\, ,
\end{equation}
where the mass scales $\Lambda_a$ are determined by the instanton actions $S_{\alpha}$, and the charge matrix $\mathbf{Q}$ has the entries
\begin{equation} \label{Qq}
	Q_a^{~i} = 2\pi\begin{pmatrix} q_\alpha^{~i} \\ q_\beta^{~i} - q_\gamma^{~i} \end{pmatrix},
\end{equation}
where $a=1,\ldots p(p+1)/2 \equiv P$.  The rows involving
$q_\beta^{~i} - q_\gamma^{~i}$ arise from cross terms in the F-term potential, see~\cite{Bachlechner:2014gfa, Long:2016jvd}.

Because the potential is periodic it is natural to define the \textit{axion fundamental domain} $\cF$~\cite{Bachlechner:2014gfa, Long:2016jvd}, given by the hyperplane constraints:
\begin{equation}
\cF = \{\theta_i \,| -\pi \leq Q_{a}^{~j}\theta_j \leq \pi\}\, .
\end{equation}
The fundamental domain
is compact when $\mathbf{Q}$ has rank $h^{1,1}$.

A quantity of key interest for axion inflation is the \emph{geometric field range}, i.e.~the maximum distance $\cR$ from the origin to the boundary of $\cF$, measured with respect to $K_{ij}$.
That is,
\begin{equation} \label{cRdef}
\cR := \max_\rho \sqrt{\mathbf{d}^{T}_\rho \cdot \mathbf{K} \cdot \mathbf{d}_\rho}\, ,
\end{equation}
where $\mathbf{d}_\rho$ is the matrix of the vertices of $\cF$, and $\mathbf{K}$ is the K\"ahler metric.  The walls and vertices of $\cF$ are determined by the integers $q_\alpha^{~i}$, i.e.~by the set of effective divisors $\{D_{\alpha}\}$ in \eqref{qwhichD} that support superpotential terms.  The problem of identifying those effective divisors of a Calabi-Yau threefold hypersurface that support nonvanishing superpotential terms has not been fully solved, cf.~\cite{Puff}.  For the purposes of the present work we will assume that every prime toric divisor $D_A$
supports a Euclidean D3-brane superpotential term, cf.~\cite{Long:2016jvd}.

Computing $\cR$ directly from \eqref{cRdef} is prohibitively expensive at large $h^{1,1}$, since the number of vertices that
must be checked is at least $2^{h^{1,1} - 1}$.
We will instead consider an upper bound on $\cR$. By performing a basis transformation
\begin{equation}\label{basistrans}
\theta_i = (\cQ^{-1})_{i}^{~j} \vartheta_j\, ,
\end{equation}
where $\mathbf{\cQ}$ is a rank $h^{1,1}$ subblock of $\mathbf{Q}$, we can trivialize $2 h^{1,1}$ of the hyperplane constraints. The metric in the $\mathbf{\vartheta}$ basis is then
\begin{equation}\label{eq:xi}
\mathbf{\Xi} = (\mathbf{\cQ}^{-1})^T \cdot \mathbf{K} \cdot  (\mathbf{\cQ}^{-1})\,,
\end{equation}
with eigenvalues $\xi_{1}^2 \le \cdots \le \xi_{h_{1,1}}^2$.
An upper bound for $\cR$ is then given by
\begin{equation}\label{eq:rmax}
\cR \leq \cR_{\rm{bound}} = \pi \sqrt{h^{1,1}} \xi_{h^{1,1}}\,,
\end{equation}
where $\xi_{h^{1,1}}^2$ is the largest eigenvalue of $\mathbf{\Xi}$. When $\mathbf{Q}$ is not square, $\cR_{\rm{bound}} $ depends on the choice of $\mathbf{\cQ}$, but each choice does provide an upper bound on $\cR$. Because we have assumed that each of the $D_A$
supports a Euclidean D3-brane superpotential term, we can
choose $h^{1,1}$ of the toric coordinates for $\mathbf{\cQ}$, in such a way that $\mathbf{\cQ}$ is the $h^{1,1} \times h^{1,1}$ identity, and $\mathbf{\Xi}=\mathbf{K}$.

\subsection{Masses}

Suppose that, for some specified K\"ahler form $J$,
and for some positive number $L$, every basis of $H_4(X,\mathbb{Z})$ contains at least $k\ge 1$ members $\Sigma_\alpha$ with $\mathrm{Vol}(\Sigma_\alpha)>L$ in string units.
Then at least $k$ axions
must have mass $\lesssim e^{-2\pi L}$.
One can therefore place upper bounds on the masses of the lightest axions by placing lower bounds on the volumes of four-cycles furnishing bases for $H_4(X,\mathbb{Z})$.

Let us first consider placing upper bounds on
superpotential contributions to axion masses, by placing lower bounds on the volumes of effective divisors.
As explained in \S\ref{inheritsec}, in this work we approximate $\mathrm{Eff}(X)$
by $\mathrm{Eff}_{\iota}(X)$; corrections to this approximation will be
described in~\cite{Picardwipmisc}.
The inherited prime toric divisors $\{D_A\}$, $A=1,\ldots,h^{1,1}+4$, provide a set of generators of $\mathrm{Eff}_{\iota}(X)$, and also, in the above approximation, of $\mathrm{Eff}(X)$.
For any $J \in H^{1,1}(X,\mathbb{R})$, not necessarily inside $\cK_X$,
we can compute the volumes
$\tau_{A}:=\frac{1}{2}\int_{D_A}J\wedge J$.  There are at most $\binom{h^{1,1}+4}{h^{1,1}}$ sets $\{D_i\}$ of $h^{1,1}$ prime toric divisors that furnish
bases for $H_4(X,\mathbb{Q})$, and for each such basis $\mathcal{B}$ we can compute the volumes $\tau_{1}^{\mathcal{B}} \le \cdots \le \tau_{h^{1,1}}^{\mathcal{B}}$ of the basis generators.  Define $\mathcal{B}_{\text{min}}$ to be the basis choice that minimizes $\tau_{h^{1,1}}^{\mathcal{B}}$.
Roughly speaking, $\mathcal{B}_{\text{min}}$ is a minimum-volume basis of generators of the effective cone.  We write
\begin{equation} \label{deftau}
\tau_{\mathrm{last}}(J) :=  \tau_{h^{1,1}}^{{\mathcal{B}}_{\text{min}}}\,,
\end{equation} denoting explicitly the dependence on the choice of $J$.
We can now give an upper bound on the magnitude of the leading superpotential term involving the lightest axion, for the given $J$:
\begin{equation}\label{wbound}
|W| \le \mathrm{exp}\bigl(-2\pi\tau_{\mathrm{last}}(J) \bigr)\,.
\end{equation}
Furthermore, given any region $\mathfrak{R} \subset H^{1,1}(X,\mathbb{R})$, not necessarily inside $\cK_X$, we can compute
\begin{equation}\label{taulastR}
\tau_{\mathrm{last}}^{\mathfrak{R}}:= \displaystyle \min_{J \in \mathfrak{R}}~\tau_{\mathrm{last}}(J)\,.
\end{equation}
We then write
\begin{equation}\label{taulastV}
\tau_{\mathrm{last}}^{V}:=\tau_{\mathrm{last}}^{\widetilde{\cK_V}}\,,
\end{equation}
\begin{equation}\label{taulastX}
\tau_{\mathrm{last}}^{X}:=\tau_{\mathrm{last}}^{\widetilde{\cK_X}}\,,
\end{equation}
\begin{equation}\label{taulastcap}
\tau_{\mathrm{last}}^{\cap}:=\tau_{\mathrm{last}}^{\widetilde{\cK_\cap}}\,.
\end{equation}
Using \eqref{inout}, we have $\tau_{\mathrm{last}}^{\cap} \le \tau_{\mathrm{last}}^{X} \le \tau_{\mathrm{last}}^{V}$.
Thus, when the condition \eqref{equ:first} for control of the $\alpha'$ expansion is imposed, the superpotential for the lightest axion is bounded above by
\begin{equation}\label{wcapis}
|W_{\cap}| :=\mathrm{exp}\left(-2\pi\tau_{\mathrm{last}}^{\cap}\right)\,.
\end{equation}
For $h^{1,1} \gg 1$ the exponentials in \eqref{axionvis} are parametrically dominant, and in evaluating the dependence of \eqref{axionvis} on the lightest axion we can omit factors that are only polynomial in the volumes, including the effect of canonical normalization.\footnote{We will verify in \S\ref{sec:results} that this is an excellent approximation, see Figure \ref{fig:mmin}.}  We then arrive at an upper bound on the mass-squared $m_{\rm{min}}^2$ of the lightest axion from \eqref{wcapis},
\begin{equation}\label{mcapis}
m_{\rm{min}}^2 \lesssim |W_{\cap}|\,.
\end{equation}
One of our main results is the computation of the bound $m_{\rm{min}}^2$ for the geometries in our ensemble.

What about axion mass terms from instanton contributions to the K\"ahler potential, resulting from Euclidean D3-branes wrapping classes $[\Sigma_\alpha] \in H_4(X,\mathbb{Z})$ that are outside $\mathrm{Eff}(X)$, and admit no holomorphic representative?  Could such instantons give masses $ \gg m_{\rm{min}}$?
We discuss this question in Appendix \ref{sec:nonholomorphic}, and find that present knowledge of minimum-volume representatives of classes outside $\mathrm{Eff}(X)$ is not sufficient to give a definite answer, but at the same time there is no evidence of such a parametric enhancement in known threefolds.  We find it plausible
that masses from $K$ are least \emph{parametrically} comparable to those from $W$, and so are approximately given by \eqref{mcapis}.

\section{Computation} \label{sec:methodology}

We obtained the topological data of Calabi-Yau threefold hypersurfaces as follows.
For each value of $h^{1,1}$ that we studied, we drew a number $\mathcal{N}(h^{1,1})$ of polytopes at random from the Kreuzer-Skarke database: see Table \ref{thetable}.\footnote{We remark in passing that the Euler number $\chi$ of $X$ is negative in more than 99\% of the geometries in our ensemble with $h^{1,1} \le 18$, but by $h^{1,1}=100$ less than 2\% of geometries have $\chi<0$.}
We manipulated the polytopes using {\tt{Sage}} \cite{sagemath}.  For each polytope that was favorable, we used {\tt{TOPCOM}} to obtain a fine and regular (but not star) triangulation.  We removed the lines in the strict interior of the polytope and included a line from the origin to each point in the polytope, thus producing an FRST $\widehat{\mathcal{T}}$~\cite{Braun:2014xka}.
Such a triangulation defines a fan, and in turn defines a toric variety $V$.
As explained in \S\ref{CYeffsec}, to study a generic Calabi-Yau threefold hypersurface,
one can omit simplices of $\widehat{\mathcal{T}}$ that pass through facets of $\Delta^{\circ}$.  We denote the set of remaining simplices by $\mathcal{T}$, and abuse language slightly in calling $\mathcal{T}$ an FRST as well.

Because we have restricted to favorable polytopes, there are $h^{1,1}+4$ prime toric divisors $\widehat{D}_A \subset V$, each corresponding to a ray of the fan determined by $\mathcal{T}$.
We picked a basis for $H_{4}(X,\mathbb{Q})$ by selecting a set of $h^{1,1}$ of the inherited prime toric divisors
$D_A := \widehat{D}_A \cap X$ that are linearly independent.
Using {\tt{Sage}}, we computed the triple intersection numbers $\kappa_{ijk}$ in the chosen basis.
Finally, we computed the Mori cone $\cM_V$ of the toric variety in {\tt Mathematica} using the algorithm described in~\cite{Berglund:1995gd}, which is equivalent to that of~\cite{oda1991}, but easier to implement.

With this data in hand, we turned to analyzing the resulting cones.  For each geometry the stretched cones $\widetilde{\cK_V}$ and $\widetilde{\cK_\cap}$ were constructed as described in \S\ref{sec:str}. We minimized the volumes $\tau_A$ and $\cV$ inside
$\widetilde{\cK_V}$ and $\widetilde{\cK_\cap}$ using {\tt{IPOPT}}, a software package for large-scale nonlinear optimization, which is included in version 11 of {\tt{Mathematica}}.
Because {\tt{IPOPT}} uses an interior point algorithm that finds a local solution to the optimization problem, we performed the minimization multiple times, from different starting points, in an attempt to find the global minimum.
Finding even one feasible starting point for the optimization algorithm is challenging at large $h^{1,1}$, as the cones $\widetilde{\cK_V}$ and $\widetilde{\cK_\cap}$ become very narrow.  We made use of IBM's optimization software {\tt{CPLEX}} as well as the {\tt{LinearProgramming}} function of {\tt{Mathematica}} to find such points.

Note that we computed \emph{one} FRST for each favorable polytope studied.  With our methods it takes of order a day
to obtain the topological data of all FRSTs of all threefolds with $h^{1,1} \le 6$, but for larger $h^{1,1}$ it quickly
becomes infeasible to compute all triangulations.  In order to provide a better point of comparison for the data we can obtain at $h^{1,1} \gg 1$, we limited ourselves to one FRST per polytope even for small $h^{1,1}$.

The values of $h^{1,1}$ that we studied, and the numbers $\mathcal{N}(h^{1,1})$, were chosen to balance the computational expense at $h^{1,1} \gg 1$ against the potential for illuminating scaling laws.  Obtaining more extensive data at large $h^{1,1}$ is an obvious next step \cite{Picardwipmisc}.  In fact, the present work has established the feasibility of obtaining the topological data of at least one threefold (i.e., one FRST) for each polytope in the Kreuzer-Skarke database.  A very rough estimate is that such a computation could require a few
CPU-centuries,
absent any improvements to the algorithms.

\begin{table}[!htp]
	\scriptsize
	\centering
	\begin{tabular}{ || c | p{4cm} | p{2.9cm} | p{3.2cm} | p{3.5cm} ||}
		\hline
		$h_{1,1}$ & \# of polytopes in KS database & \# of polytopes studied & \# of favorable polytopes & \# of volume minimizations \\
		\hline
		2 & 36 & 36 & 36 & 36 \\
		
		3 & 244 & 244 & 243 & 243\\
		
		4 & 1197 & 1,197 & 1,185 & 1,185 \\
		
		5 & 4,990 & 4,990 & 4,987 & 3,000 \\
		
		6 & 17,101 & 17,101 & 16,608 & 3,000\\
		
		7 & 50,376 & 50,376 & 48,221& 3,000\\
		
		8 & 128,165 & 128,165 & 120,759& 3,000\\
		
		9 & 285,929 & 285,929 & 264,558 & 3,000\\
		
		10 & 568,078 & 568,078 & 515,319 & 3,000\\
		
		11 & 1,022,264 & 300,000 & 261,541 & 3,000\\
		
		12 & 1,685,784 & 100,000 & 86,860 & 3,000\\
		
		13 & 2,580,222 & 100,000 & 84,923 & 3,000\\
		
		14 & 3,697,767 & 100,000 & 82,939 & 3,000\\
		
		15 & 5,011,933 & 100,000 & 80,415 & 3,000\\
		
		16 & 6,473,431 & 100,000 & 78,756 & 3,000\\
		
		17 & 7,989,780 & 100,000 & 76,749 & 3,000\\
		
		18 & 9,561,562 & 100,000 & 75,109 & 3,000\\
		
		19 & 11,054,578 & 100,000 & 73,454 & 3,000\\
		
		20 & 12,434,427 & 100,000 & 71,656 & 3,000\\
		
		21 & 13,652,664 & 20,000 & 14,136 & 3,000\\
		
		22 & 14,677,475 & 20,000 & 13,844 & 3,000\\
		
		23 & 15,484,811 & 3,000 & 2,047 & 2,047\\
		
		24 & 16,088,119 & 3,000 & 2,025 & 2,025\\
		
		25 & 16,495,690 & 3,000 & 1,988 & 1,988\\
		
		30 & 15,914,795 & 3,000 & 1,907 & 1,907\\
		
		35 & 12,955,936 & 3,000 & 1,866 & 1,866\\
		
		40 & 9,620,216 & 3,000 & 1,808 & 1,808\\
		
		45 & 6,787,275 & 3,000 & 1,774 & 1,774\\
		
		50 & 4,659,208 & 3,000 & 1,729 & 1,729\\
		
		55 & 3,171,468 & 3,000 & 1,700 & 1,700\\
		
		60 & 2,174,347 & 3,000 & 1,654 & 1,654\\
		
		65 & 1,494,731 & 3,000 & 1,634 & 1,634\\
		
		70 & 1,018,865 & 3,000 & 1,641 & 1,641\\
		
		75 & 762,815 & 3,000 & 1,627 & 1,627\\
		
		80 & 487,805 & 3,000 & 1,655 & 1,655\\
		
		85 & 339,574 & 3,000 & 1,641 & 1,641\\
		
		90 & 246,570 & 3,000 & 1,604 & 1,604\\
		
		95 & 179,981 & 3,000 & 1,629 & 1,629\\
		
		100 & 129,605 & 3,000 & 1,626 & 1,626\\	
		
		105 & 92,887 & 3,000 & 1,597 & 0 \\
		
		110 & 68,453 & 3,000 & 1,627 & 0 \\
		
		115 & 51,509 & 3,000 & 1,619& 0 \\
		
		120 & 39,847 & 3,000 & 1,602& 0 \\
		
		130 & 23,001 & 3,000 & 1,597& 0 \\
		
		135 & 16,731 & 3,000 & 1,659& 0 \\
		
		140 & 12,392 & 3,000 & 1,626& 0 \\
		
		145 & 9,411 & 3,000 & 1,596& 0 \\
		
		155 & 5,440 & 3,000 & 1,646& 0 \\
		
		160 & 4,101 & 3,000 & 1,697& 0 \\
		
		165 & 3,160 & 3,000 & 1,717& 0 \\
		
		170 & 2,502 & 2502 & 1,403& 0 \\
		
		180 & 1,486 & 1486 & 899& 0 \\
		
		185 & 1,318 & 1318 & 750& 0 \\
		
		190 & 1,209 & 1209 & 685& 0 \\
		
		195 & 830 & 830 & 497& 0 \\
		
		205 & 535 & 535 & 324& 0 \\
		
		210 & 483 & 483 & 276& 0 \\
		
		215 & 392 & 392 & 233& 0 \\
		
		220 & 356 & 356 & 208& 0 \\
		
		230 & 219 & 219 & 113& 0 \\
		
		235 & 172 & 172 & 113& 0 \\
		
		240-491 & 4,358 & 4,358 & 2,671& 0 \\
		
		\hline
		\hline
	\end{tabular}
	\caption{The dataset.}  \label{thetable}
\end{table}

\FloatBarrier

\section{Results} \label{sec:results}

The primary topological data produced by our analysis are the generators of the Mori cones $\cM_V$ of toric varieties $V$, and the intersection numbers $\kappa_{ABC}$ of inherited prime toric divisors $D_A$ of Calabi-Yau hypersurfaces $X \subset V$.  Taking these data and imposing the condition \eqref{equ:first}, we can compute the stretched K\"ahler cones $\widetilde{\cK_V}$ and $\widetilde{\cK_\cap}$, which bound the stretched K\"ahler cone
$\widetilde{\cK_X}$ of $X$ from the inside and the outside, respectively, cf.~\eqref{inout}.  Then, for any holomorphic $2k$-cycle $\Sigma_{2k}$ ($1\le k \le 3$) in $X$, $\widetilde{\cK_\cap}$ determines a lower bound on $\mathrm{Vol}(\Sigma_{2k}) \equiv \frac{1}{k!}\int_{\Sigma_{2k}} \wedge^k J$.

In this section we report salient features of the intersection numbers, Mori cones,
volumes of holomorphic cycles, geometric field ranges, and
masses in our ensemble.

\subsection*{Topological Data:}

The volumes $\cV$, $\tau_A$ and $\mathfrak{t}_{AB}$ depend on the intersection numbers $\kappa_{ijk}$, as given in \eqref{eq:thevols}.  Since $\kappa_{ijk}$ depends on a choice of basis of $H_4(X,\mathbb{Z})$,
we instead report the statistical properties of $\kappa_{ABC}$, which is basis-independent.

\begin{figure}
	\centering
	\begin{minipage}{1\textwidth}
		\centering
		\includegraphics[width=9cm]{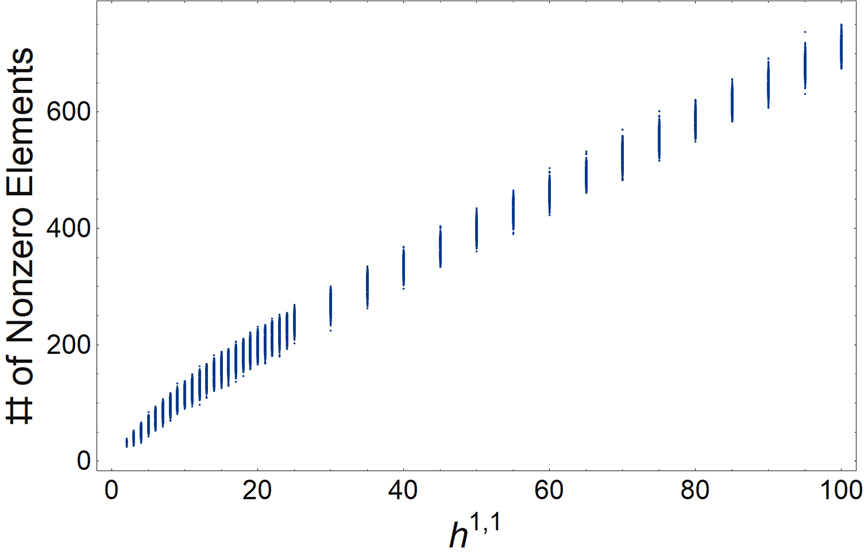}
		\caption{Number of nonzero entries of $\kappa_{ABC}$, cf.~\eqref{intnumberdef}, vs.~$h^{1,1}$.}
		\label{fig:nonzero}
	\end{minipage}\vspace{10.00mm}
	\begin{minipage}{1\textwidth}
		\centering
		\includegraphics[width=9cm]{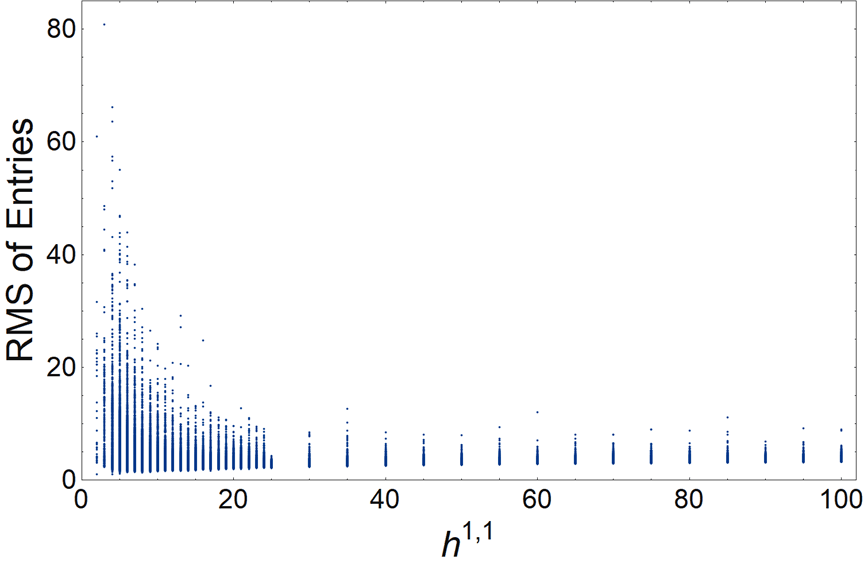}
		\caption{Root mean square size of nonzero entries of $\kappa_{ABC}$ vs.~$h^{1,1}$.}
		\label{fig:rms}
	\end{minipage}\vspace{10.00mm}
	\begin{minipage}{1\textwidth}
		\centering
		\includegraphics[width=9cm]{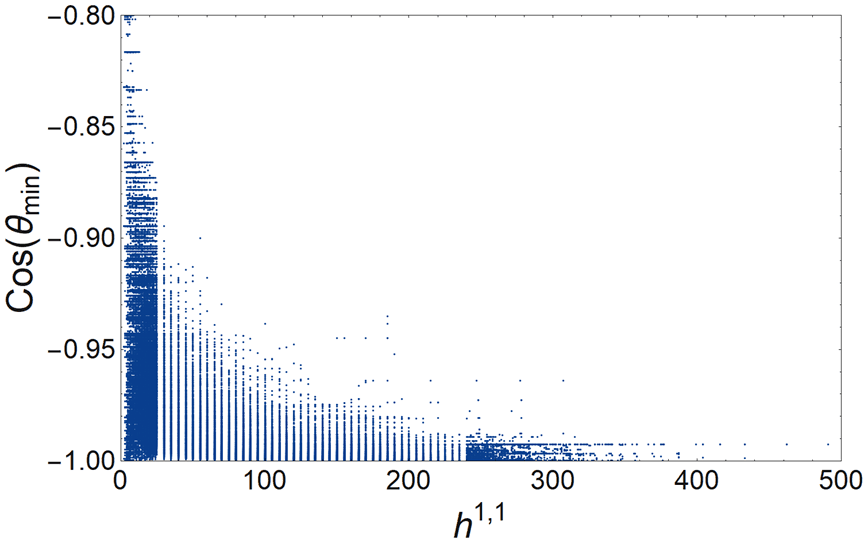}
		\caption{$\cos (\theta_{\text{min}})$, defined in (\ref{eq:minang}), vs.~$h^{1,1}$.}
		\label{fig:minang}
	\end{minipage}
\end{figure}

We first examine the sparseness of $\kappa_{ABC}$. The number of nonvanishing intersection numbers per geometry increases approximately linearly with $h^{1,1}$, as shown in Figure~\ref{fig:nonzero}. As a result, $\kappa_{ABC}$ becomes very sparse at large $h^{1,1}$.
In Figure~\ref{fig:rms} we show the root mean square (RMS) size of the nonvanishing intersection numbers for each geometry.

The cone $\cK_V$ is given by the intersection of the half-spaces defined by the linear inequalities
\begin{equation}
	M_{ai}t^i >0.
\end{equation}
As $h^{1,1}$ increases, the number of inequalities grows and $\cK_V$ becomes very narrow.
A conceptually straightforward way to quantify the narrowness of the cone $\cK_V$
would be to analyze the behavior of the solid angle subtended by $\cK_V$ as a function of $h^{1,1}$. However, this becomes computationally expensive when $h^{1,1} \gtrsim 15$. Instead, we characterize the narrowness of $\cK_V$ by computing the cosine of the smallest angle between two hyperplanes, denoted $M^a$ and $M^b$:
\begin{equation}\label{eq:minang}
	\cos (\theta_{\text{min}}) := \displaystyle\min_{a,b} \Biggl(\frac{M^a \cdot M^b}{|M^a||M^b|}\Biggr)\,.
\end{equation}
As the angle $\theta_\text{min}$ between two hyperplanes approaches zero, the cone becomes infinitely narrow. This can also be understood from the perspective of the dual cone $\cM_V$. When $\cM_V$ has two generating rays $M^a$ and $M^b$ that are almost antiparallel (such that $\cK_V$ has facets whose normals are almost antiparallel), it is difficult to find a K\"ahler form $J$ such that both of the associated curves have positive volumes simultaneously. Figure \ref{fig:minang} shows $\cos (\theta_\text{min})$ as a function of $h^{1,1}$.

\FloatBarrier

\begin{figure}
	\centering
	\begin{subfigure}[t]{1\textwidth}
		\centering
		\includegraphics[width=14cm]{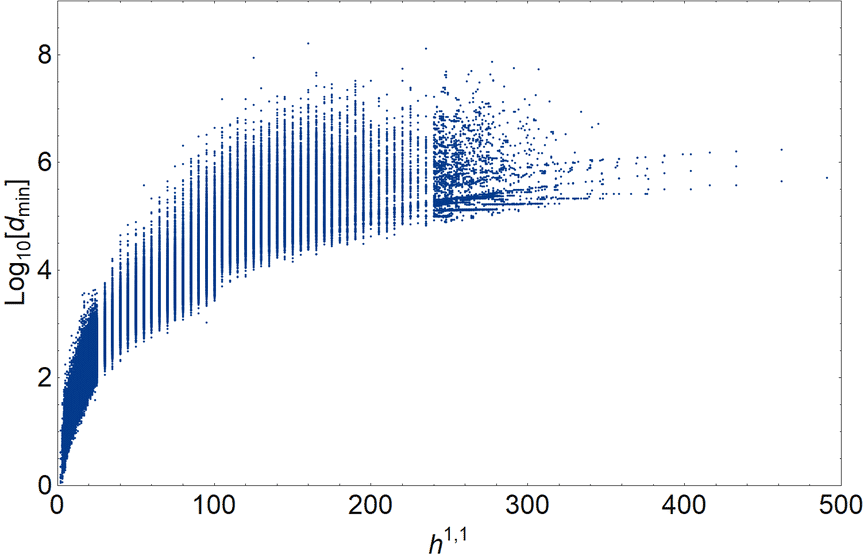}
	\end{subfigure}\vspace{.5cm}
	\begin{subfigure}[t]{1\textwidth}
		\centering
		\includegraphics[width=14cm]{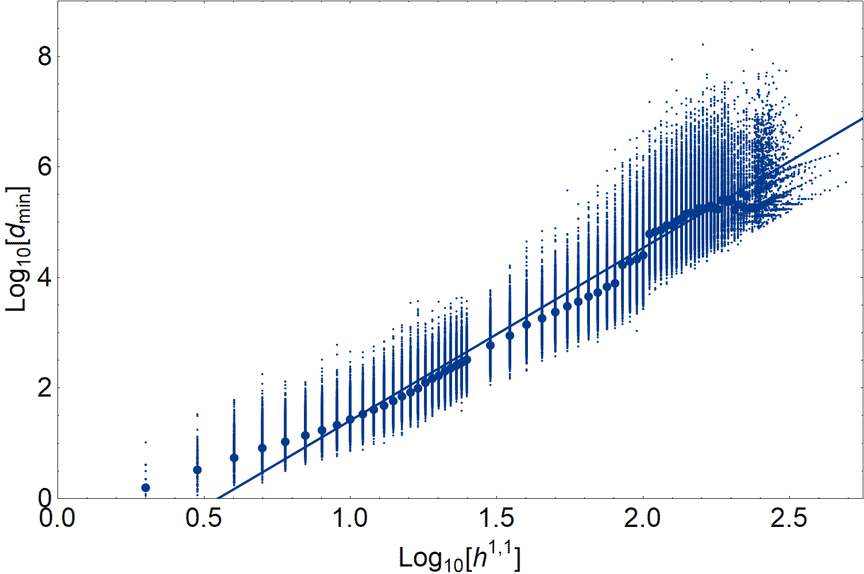}
	\end{subfigure}\vspace{.5cm}
	\caption{$\log_{10}(\text{d}_{\text{min}}^V)$, defined in \eqref{eq:dminV}, vs.~$h^{1,1}$ and vs.~$\log_{10}(h^{1,1})$. The fit is $\log_{10}(\text{d}_{\text{min}}^V)= -1.7 + 3.1 \log_{10} (h^{1,1})$.}
	\label{fig:dminKV}
\end{figure}

\FloatBarrier

\begin{figure}[ht]
	\centering
	\includegraphics[width=9cm]{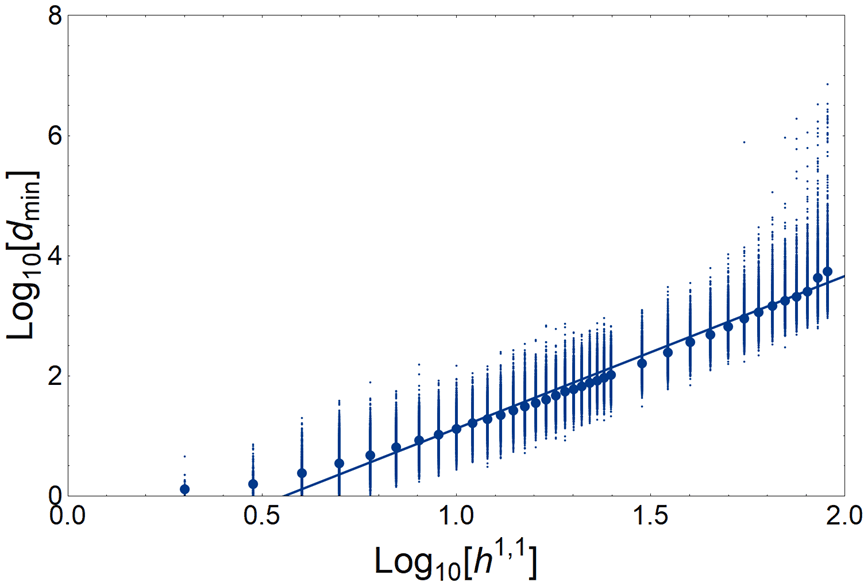}
	\caption{$\log_{10}(\text{d}_{\text{min}}^\cap)$, defined in \eqref{eq:dminint}, vs.~$\log_{10}(h^{1,1})$. The fit is $\log_{10}(\text{d}_{\text{min}}^\cap)= -1.4 + 2.5 \log_{10}(h^{1,1})$.}
	\label{fig:dminKint}
\end{figure}

As $\cK_V$ becomes more narrow, the stretched cone $\widetilde{\cK_V}$, defined in \eqref{eq:stretch2}, gets pushed further away from the origin. Another measure of the size of $\cK_V$ is therefore the shortest distance $d_{\text{min}}^V$ between the origin and any point of $\widetilde{\cK_V}$,
\begin{equation}\label{eq:dminV}
d_{\text{min}}^V := \displaystyle \min_{t^i} \Big\{\sqrt{t^i t_i} \,\Big|\, t^i[D_i] \in \widetilde{\cK_V}\Big\}\,,
\end{equation}
and we denote the minimum-distance point by $t_{d}^{V}$.  See Figure \ref{fig:dminKV}.

Although $\cK_V$ is computationally accessible (even for $h^{1,1}=491$), and the size of $\cK_V$ is generally correlated with the size of $\cK_X$, $\cK_V$ can in principle be much more narrow than $\cK_X$.  Analysis of $\cK_V$ alone can therefore provide only estimates of the volumes of holomorphic cycles in $X$, for $t^i [D_i] \in \cK_X$, rather than definite bounds.  To obtain lower bounds on cycle volumes, we instead examine $\cK_\cap$, which $\textit{contains}$ $\cK_X$. The tradeoff is that $\cK_\cap$ is a complicated cone defined by linear, quadratic and cubic constraints, and defining a quantity analogous to $\theta_{\mathrm{min}}$ is difficult. We can, however, compute $\text{d}_{\text{min}}^\cap$, the shortest distance between the origin and any point of $\widetilde{\cK_\cap}$,
\begin{equation}\label{eq:dminint}
d_{\text{min}}^\cap := \displaystyle \min_{t^i}
\Big\{\sqrt{t^i t_i} \,\Big|\, t^i[D_i] \in \widetilde{\cK_\cap}\Big\}\,,
\end{equation}
and we denote the minimum-distance point by $t_{d}^{\cap}$.  See Figure \ref{fig:dminKint}.

\subsection*{Volumes:}
\begin{figure}[ht]
	\centering
	\begin{subfigure}[t]{1\textwidth}
		\centering
		\includegraphics[width=9cm]{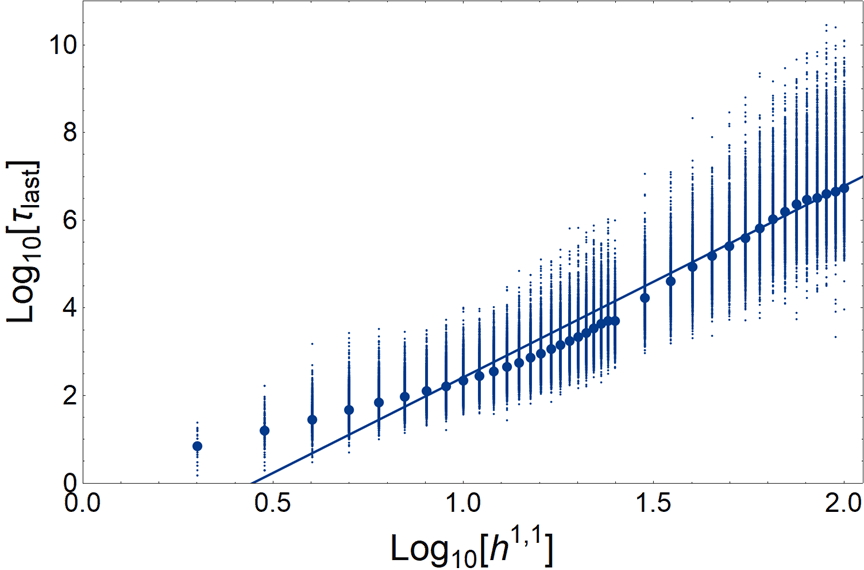}
		\caption{$\log_{10} (\tau_{\text{last}}^V)$ vs.~$\log_{10}(h^{1,1})$. The fit is $\log_{10} (\tau_{\text{last}}^V)= -1.9 + 4.3 \log_{10}(h^{1,1})$.}
	\end{subfigure}\vspace{.5cm}
	\begin{subfigure}[t]{1\textwidth}
		\centering
		\includegraphics[width=9cm]{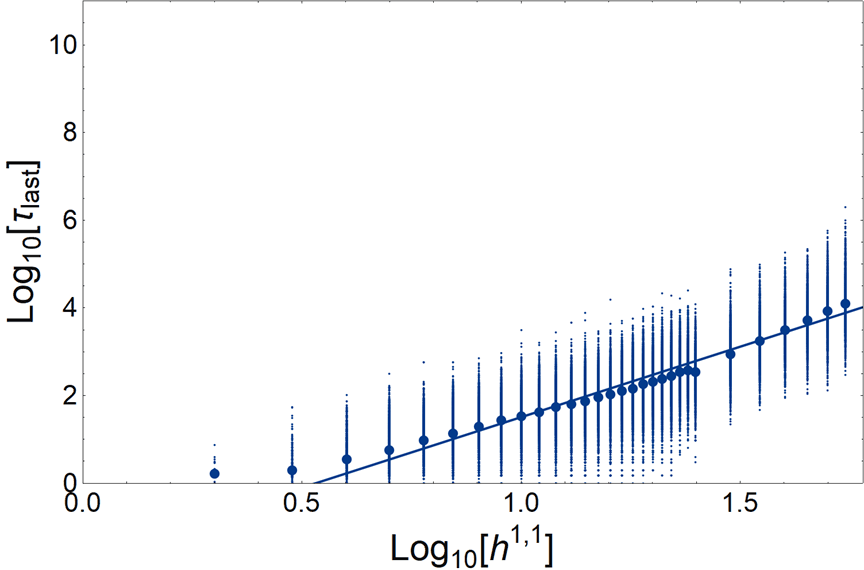}
		\caption{$\log_{10}(\tau_{\text{last}}^\cap)$ vs.~$\log_{10}(h^{1,1})$. The fit is $\log_{10} (\tau_{\text{last}}^\cap)= -1.7 + 3.2 \log_{10}(h^{1,1})$.}
	\end{subfigure}
	\caption{Lower bounds on $\tau_{\text{last}}^V$, defined in \eqref{taulastV}, and $\tau_{\text{last}}^\cap$, defined in \eqref{taulastcap}, vs.~$h^{1,1}$.}
	\label{fig:divvol}

\vspace{0.6cm}

	\begin{subfigure}[t]{0.45\textwidth}
		\centering
		\includegraphics[width=1\linewidth]{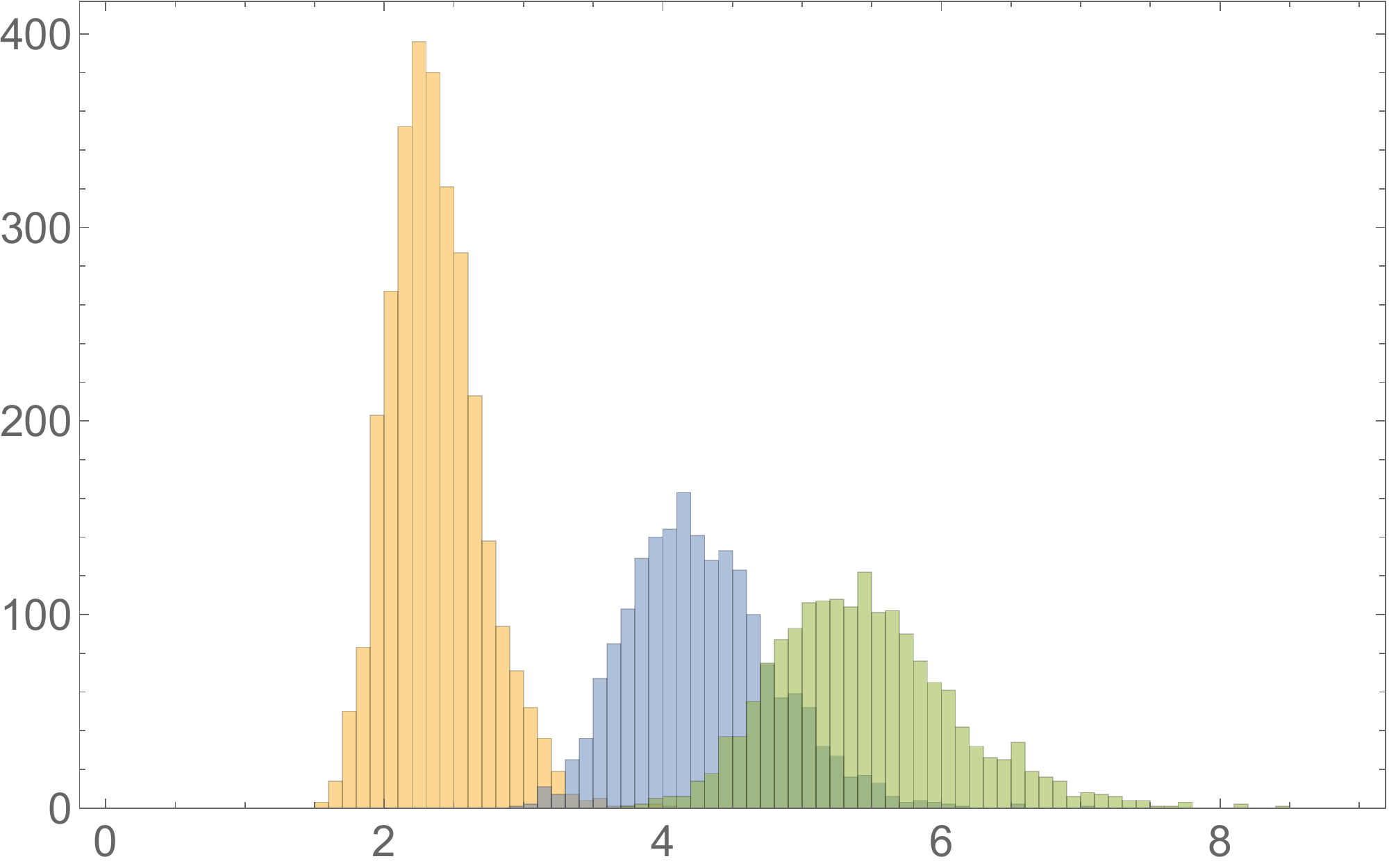}
		\caption{$\log_{10} (\tau_{\text{last}}^V)$.}
	\end{subfigure}\hfill
	\begin{subfigure}[t]{0.45\textwidth}
		\centering
		\includegraphics[width=1\linewidth]{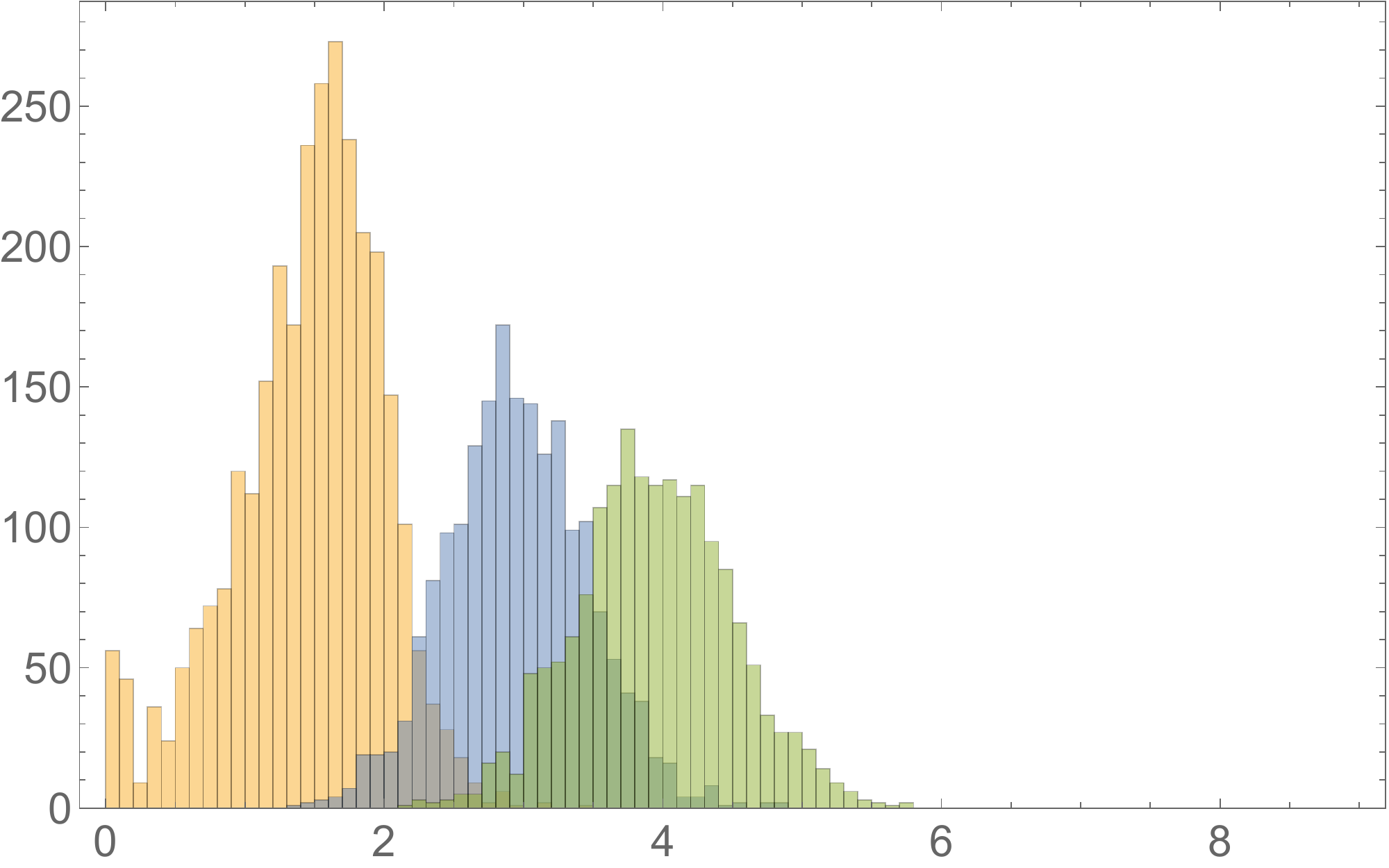}
		\caption{$\log_{10} (\tau_{\text{last}}^\cap)$.}
	\end{subfigure}\hfill
	\caption{$\tau_{\text{last}}^V$ (left) and $\tau_{\text{last}}^\cap$ (right) for $h^{1,1}=10$ (leftmost peak), $30$ (center peak), and $50$ (rightmost peak).}
	\label{fig:divvolhist}
\end{figure}

\begin{figure}[ht]
	\centering
	\begin{subfigure}[t]{1\textwidth}
		\centering
		\includegraphics[width=9cm]{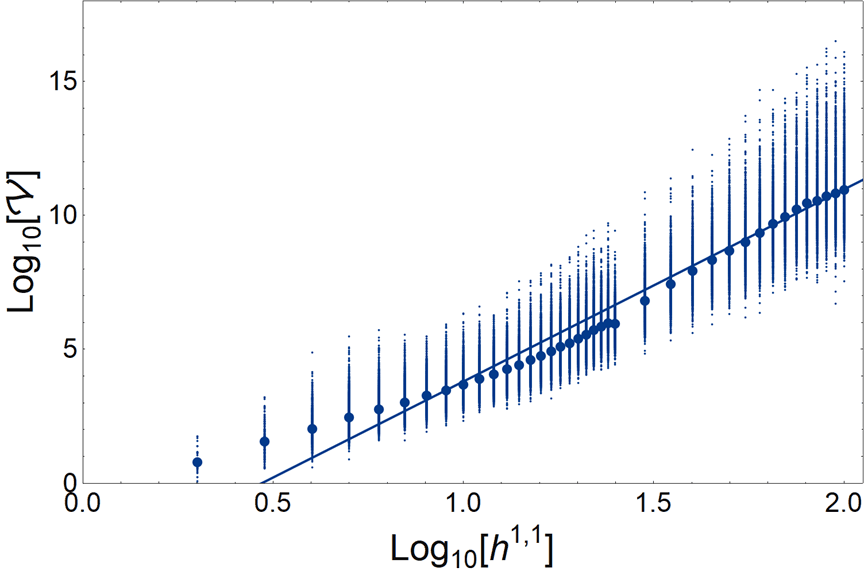}
		\caption{$\log_{10} (\cV^V)$ vs.~$\log_{10}(h^{1,1})$. The fit is $\log_{10} (\cV^V)= -3.4 + 7.2 \log_{10}(h^{1,1})$.}
	\end{subfigure}\vspace{.5cm}
	\begin{subfigure}[t]{1\textwidth}
		\centering
		\includegraphics[width=9cm]{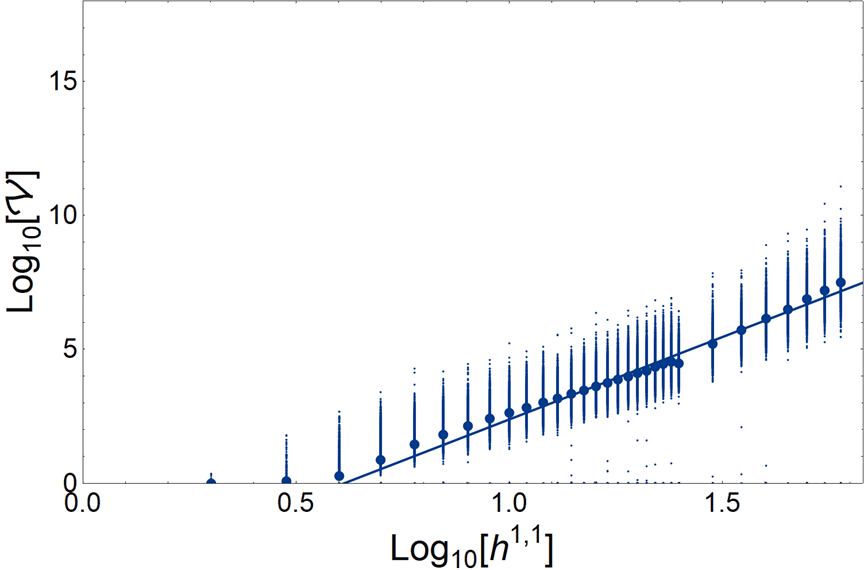}
		\caption{$\log_{10} (\cV^\cap)$ vs.~$\log_{10}(h^{1,1})$. The fit is $\log_{10} (\cV^\cap)= -3.8 + 6.2 \log_{10}(h^{1,1})$.}
	\end{subfigure}\hfill
	\caption{Lower bounds on $\cV$, defined in \eqref{eq:thevols}, vs.~$h^{1,1}$ in $\widetilde{\cK_V}$ (top) and $\widetilde{\cK_\cap}$ (bottom).}
	\label{fig:cyvol}

\vspace{0.6cm}

	\begin{subfigure}[t]{0.45\textwidth}
		\centering
		\includegraphics[width=1\linewidth]{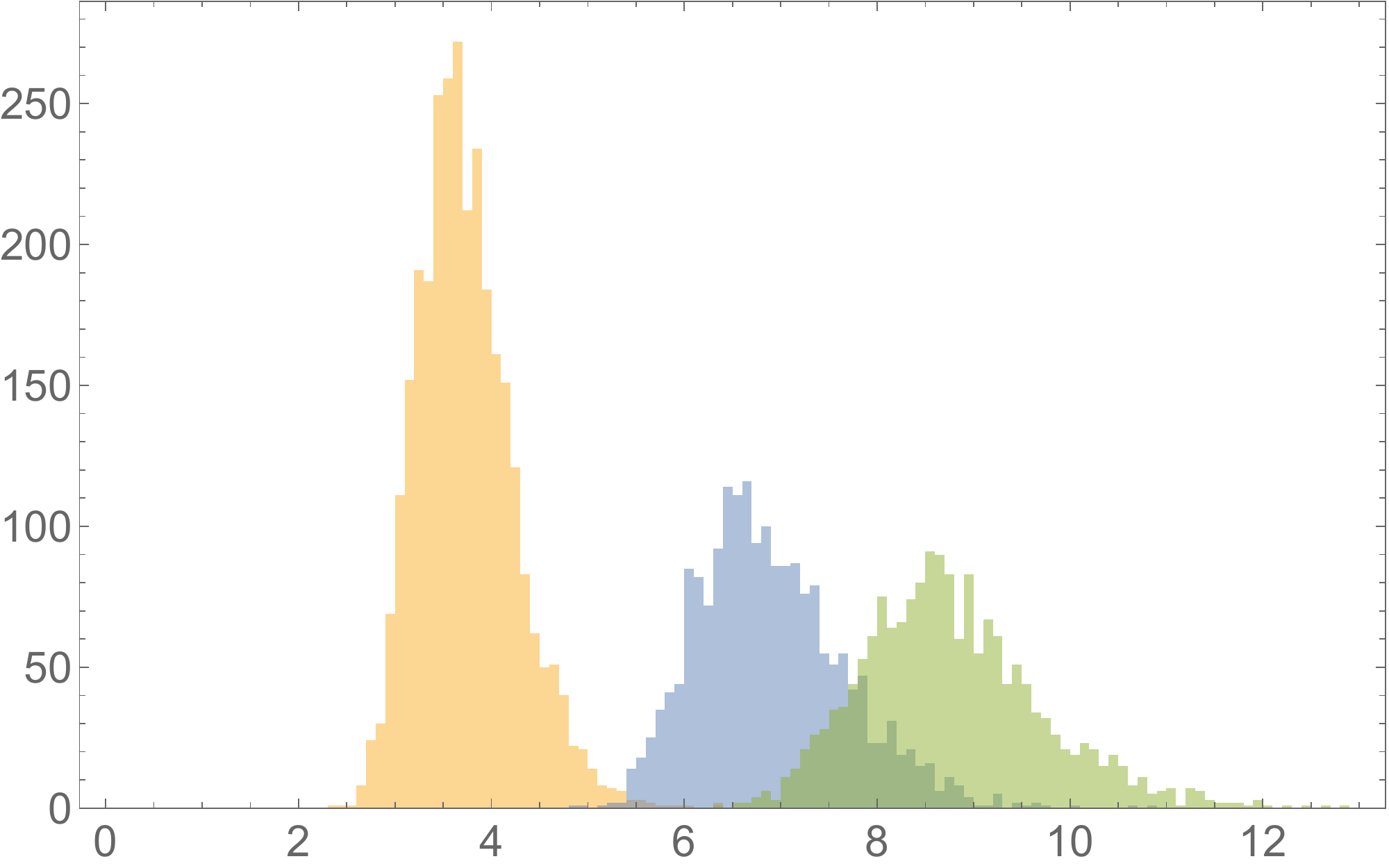}
		\caption{$\log_{10} (\cV^V)$.}
	\end{subfigure}\hfill
	\begin{subfigure}[t]{0.45\textwidth}
		\centering
		\includegraphics[width=1\linewidth]{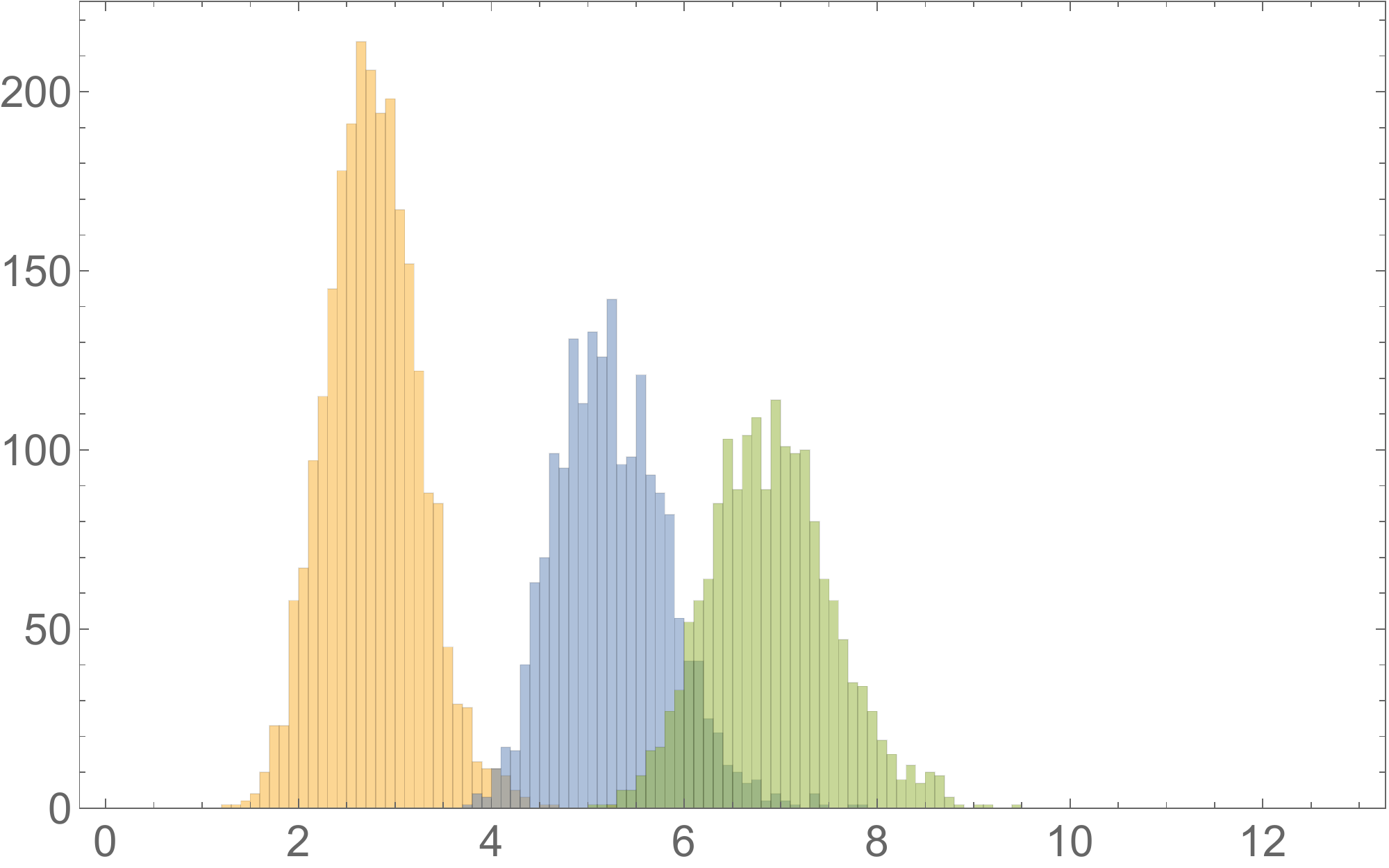}
		\caption{$\log_{10} (\cV^\cap)$.}
	\end{subfigure}\hfill
	\caption{Lower bounds on $\cV$ in $\widetilde{\cK_V}$ (left) and $\widetilde{\cK_\cap}$ (right) for $h^{1,1}=10$ (leftmost peak), $30$ (center peak), and $50$ (rightmost peak).}
	\label{fig:cyvolhist}
\end{figure}

To compute lower bounds on $\tau_{\text{last}}$ and $\cV$, for each prime toric divisor $D_A$ we
numerically minimize the divisor volume $\tau_A$ in $\widetilde{\cK_V}$ and in $\widetilde{\cK_\cap}$.  We then calculate
$\tau_{\text{last}}^{V}$  and $\tau_{\text{last}}^{\cap}$
as described in \S4 and \S5. The resulting bounds are shown in Figures \ref{fig:divvol}-\ref{fig:cyvolhist}.\footnote{We omit cases in which the only $\cK_\cap$ constraint on $\mathcal{V}$ is the trivial one $\mathcal{V}>1$, cf.~\eqref{defkcap}: for these geometries a direct computation of $\cK_X$ is plausibly necessary.}

\newpage

\FloatBarrier

\subsection*{Geometric field ranges:}

\begin{figure}[ht]
	\centering
	\begin{subfigure}[t]{0.45\textwidth}
		\centering
		\includegraphics[width=1\linewidth]{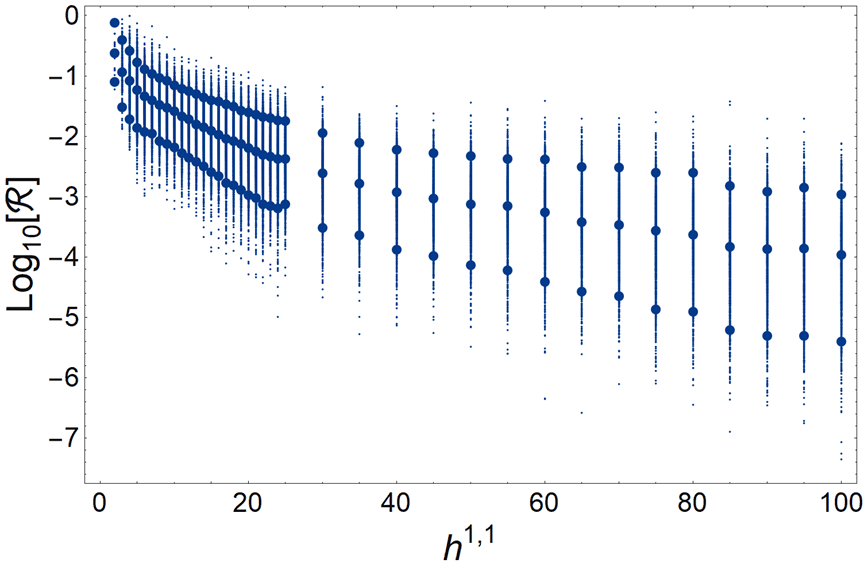}
		\caption{$\cR_{\text{bound}}^V$ vs.~$h^{1,1}$.}
	\end{subfigure}\hfill
	\begin{subfigure}[t]{0.45\textwidth}
		\centering
		\includegraphics[width=1\linewidth]{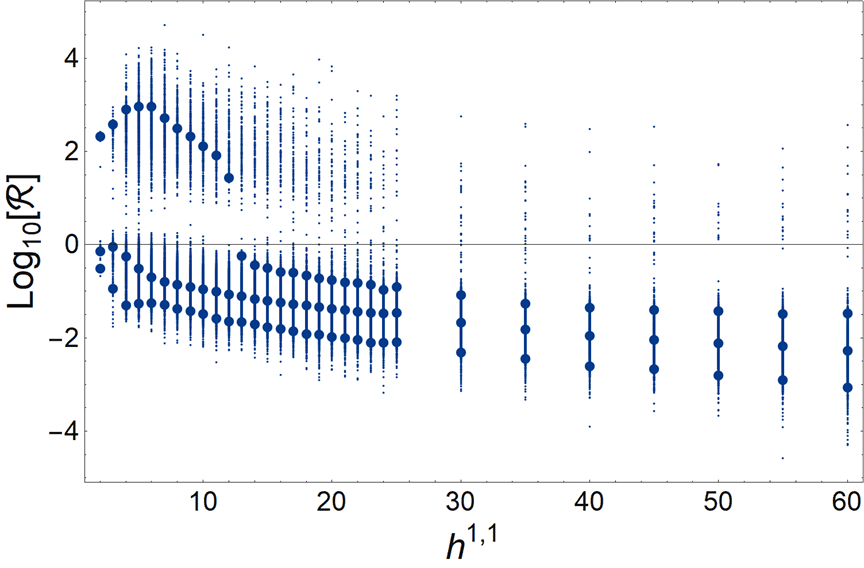}
		\caption{$\cR_{\text{bound}}^\cap$ vs.~$h^{1,1}$.}
	\end{subfigure}\hfill
	\caption{Upper bounds on the geometric field range, cf.~\eqref{crVis}, vs.~$h^{1,1}$.  Left: $\log_{10}(\cR_{\text{bound}}^V)$.  Right: $\log_{10}(\cR_{\text{bound}}^\cap)$.  5th, 50th, and 95th percentiles are shown.}
	\label{fig:axrad}
\end{figure}

As explained in \S\ref{sec:masses},
we estimate the radius $\cR$ \eqref{cRdef} of the axion fundamental domain $\cF$  by assuming that all
prime toric divisors $D_A$
contribute to the superpotential.
The radius depends on the K\"ahler parameters $t^i$, and we report upper bounds on $\cR$ at two locations.
We define $t_{\mathcal{V}}^{V}$ and $t_{\mathcal{V}}^{\cap}$ to be the points in  $\widetilde{\cK_V}$ and  $\widetilde{\cK_\cap}$, respectively, where the threefold volume $\cV$ is minimized, and we define
\begin{equation}\label{crVis}
\cR^{V} :=  \cR(t_{\mathcal{V}}^{V}), \qquad  \cR^{\cap} :=  \cR(t_{\mathcal{V}}^{\cap})\,.
\end{equation}
We first compute the K\"ahler metric $K_{ij}$ at $t_{\mathcal{V}}^{V}$.
We next trivialize $2h^{1,1}$ of the hyperplane constraints, as in \eqref{basistrans}, taking $\mathbf{\cQ}$ to be the $h^{1,1} \times h^{1,1}$ identity subblock of the charge matrix $\mathbf{\cQ}$ corresponding to a choice of $h^{1,1}$ of the toric coordinates.  This yields an upper bound $\cR^V_{\rm{bound}} \ge \cR^V$, shown
in Figure \ref{fig:axrad}.  Computing $K_{ij}$ instead at $t_{\mathcal{V}}^{\cap}$ and repeating the trivialization, we obtain the upper bound $\cR^\cap_{\rm{bound}} \ge \cR^\cap$, shown in Figure \ref{fig:axrad}.

\subsubsection*{Axion masses:}

\begin{figure}[ht]
	\centering
	\begin{subfigure}[t]{0.45\textwidth}
		\centering
		\includegraphics[width=7cm]{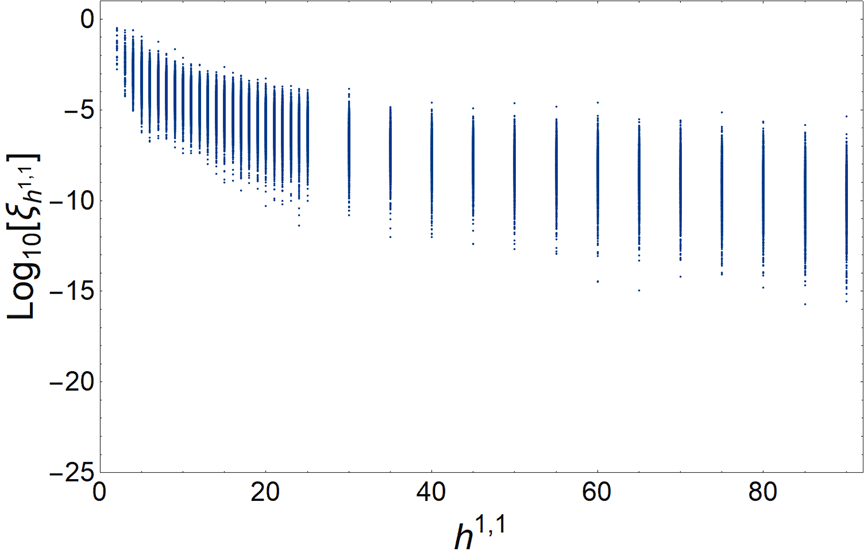}
		\caption{$\log_{10} (\xi^V_{h^{1,1}})$ vs.~$h^{1,1}$.}
	\end{subfigure}\hfill
	\begin{subfigure}[t]{0.45\textwidth}
		\centering
		\includegraphics[width=7cm]{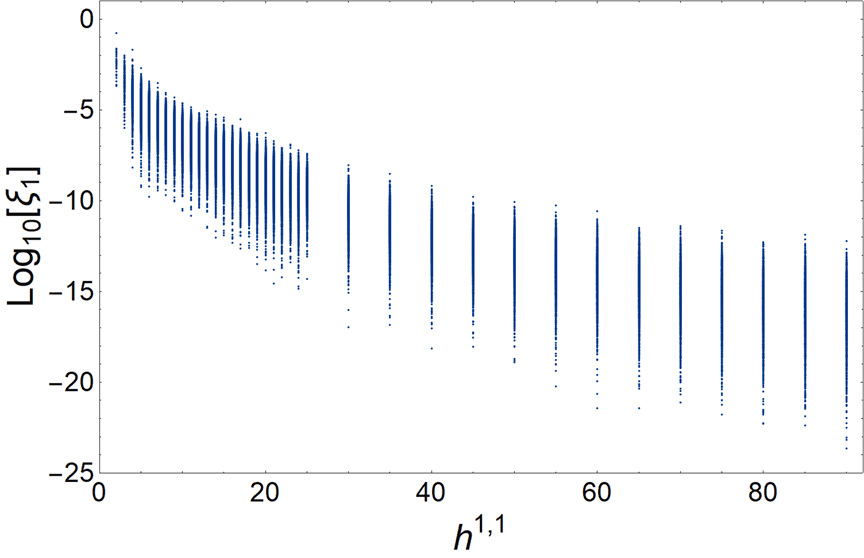}
		\caption{$\log_{10} (\xi^V_{1})$ vs.~$h^{1,1}$.}
	\end{subfigure}\vspace{.8cm}
	\begin{subfigure}[t]{0.45\textwidth}
		\centering
		\includegraphics[width=7cm]{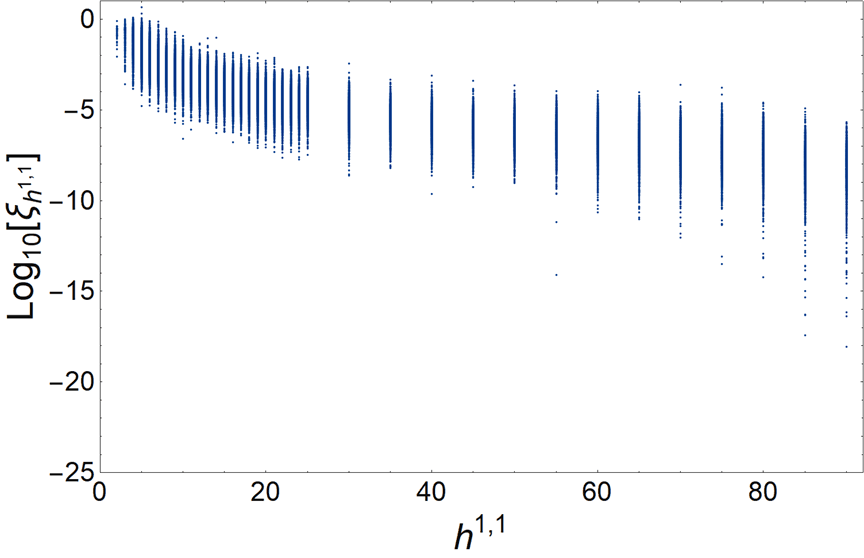}
		\caption{$\log_{10} (\xi^\cap_{h^{1,1}})$ vs.~$h^{1,1}$.}
	\end{subfigure}\hfill
	\begin{subfigure}[t]{0.45\textwidth}
		\centering
		\includegraphics[width=7cm]{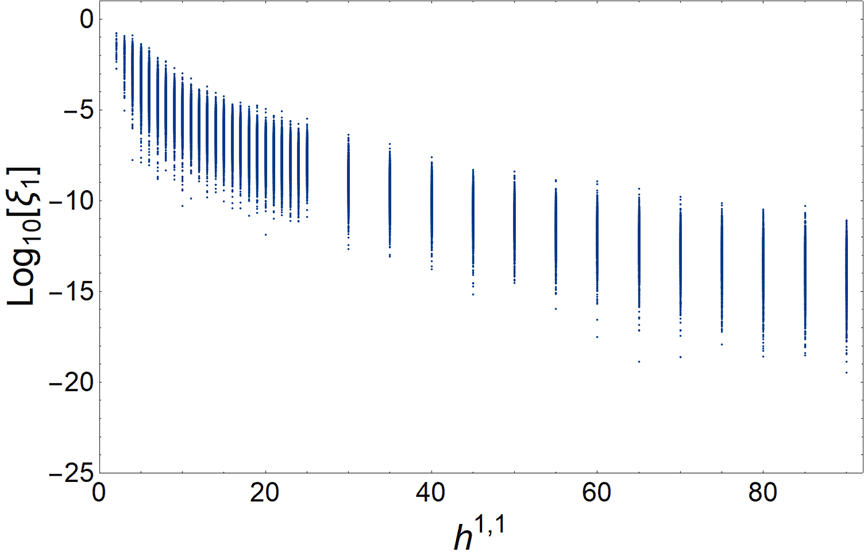}
		\caption{$\log_{10} (\xi^\cap_{1})$ vs.~$h^{1,1}$.}
	\end{subfigure}\hfill
	\caption{Maximum (left) and minimum (right) eigenvalues of the kinetic matrix $\Xi$, defined in \eqref{eq:xi}, vs.~$h^{1,1}$, evaluated at $t_{\mathcal{V}}^{V}$ (top) and $t_{\mathcal{V}}^{\cap}$ (bottom).}
	\label{fig:keigval}
\end{figure}

\begin{figure} \centering
	\includegraphics[width=11cm]{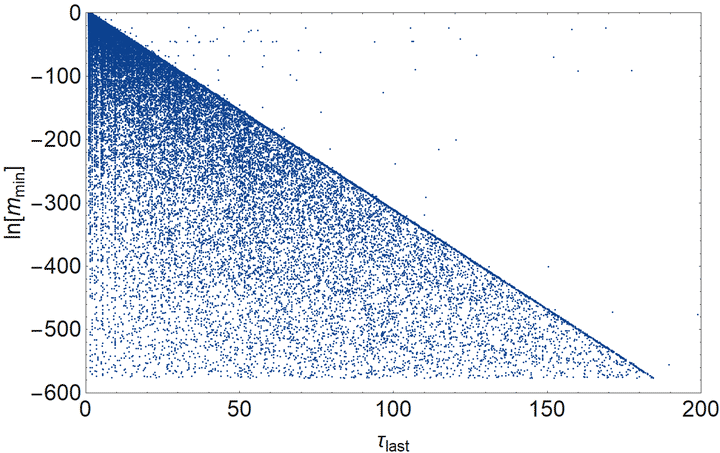}
	\caption{$\mathrm{ln} (m_\text{min})$ evaluated at $t_d^{\cap}$, cf.~\eqref{mmindef}, vs.~$\tau_{\text{last}}^\cap$. The edge is at
		$\mathrm{ln} (m_{\text{min}}) = - \pi \tau_{\text{last}}$.  }
	\label{fig:mmin}
\end{figure}

Now consider type IIB string theory compactified on an orientifold of a hypersurface $X$ from our ensemble.
The large divisor volumes lead to powerful suppression of superpotential contributions to the potential for $C_4$ axions $\theta_i$.
We find that in
every geometry in our ensemble\footnote{One must bear in mind that we have examined a very limited sample of the Kreuzer-Skarke list, and so our findings should be understood as indicating typical behavior, not establishing a no-go.} with $h^{1,1} > 22$, the lightest axion is essentially massless, with the canonically-normalized field having mass
\begin{equation}\label{mminceiling}
m < 10^{-33}\,\rm{eV}\,.
\end{equation}

Let us also give a heuristic estimate of the expected mass of the lightest axion.
By the definition \eqref{taulastR}, every basis for $H_4(X,\mathbb{Q})$ consisting of elements of $\mathrm{Eff}(X)$ has members with volume $\ge \tau_{\rm{last}}^{\cap}$.
As seen from the fit in Figure \ref{fig:divvol}, $\tau_{\rm{last}}^{\cap} \sim 0.02(h^{1,1})^p$ with $p\sim 3$.
Hence, one or more of the $h^{1,1}$ axions $\theta_i$ receives no superpotential contributions larger than
\begin{equation}\label{wcapisres}
|W_{\cap}| \equiv \mathrm{exp}(-2\pi\tau_{\rm{last}}^{\cap}) \sim \mathrm{exp}\Bigl(-0.1(h^{1,1})^3 \Bigr)\,.
\end{equation}
The exponent $p$ changes, within the range $3 \lesssim p \lesssim 6$, depending on whether one examines $\tau_{\rm{last}}^{\cap}$ --- which is the most direct and conservative --- or instead a more computable proxy such as $\tau_{\rm{last}}^{V}$ or $(d_{\text{min}}^V)^2$.
However, such changes do not alter our central finding that one or more axions are extremely light when $h^{1,1} \gg 1$ and $J \in \widetilde{\cK_\cap}$.

\subsection*{Summary:}

A root cause of our findings is that the K\"ahler cones of Calabi-Yau threefold hypersurfaces are very narrow for $h^{1,1} \gg 1$, as shown in Figure \ref{fig:minang}.
The condition \eqref{equ:first} that every effective curve has volume $\ge 1$, which we have used as a proxy for control of the $\alpha'$ expansion, then implies that the  K\"ahler form $J \in H^{1,1}(X,\mathbb{R})$ is far from the origin in $H^{1,1}(X,\mathbb{R})$, in the sense of \eqref{eq:dminV}: see Figures \ref{fig:dminKV} and \ref{fig:dminKint}.
In turn, many irreducible effective curves and irreducible effective divisors have large volumes, see Figures \ref{fig:divvol}-\ref{fig:divvolhist}.
Furthermore, the volume $\mathcal{V}$ of $X$ itself is large (Figures \ref{fig:cyvol}-\ref{fig:cyvolhist}), the geometric field range is generally small (Figure \ref{fig:axrad}),
and the eigenvalues of the axion kinetic matrix are small (Figure \ref{fig:keigval}).  The minimum axion mass is small, and strongly correlated with $\tau_{\rm{last}}^{\cap}$ (Figure \ref{fig:mmin}).\footnote{In fact, $m_\text{min}(t_d^{\cap})$ is almost perfectly correlated with $\tau_{\rm{last}}(t_d^{\cap})$.
Note that by \eqref{deftau}, $\tau_{\rm{last}}^{\cap} \neq \tau_{\rm{last}}(t_d^{\cap})$.}

\FloatBarrier

\section{Implications for Axion Cosmology} \label{sec:implications}

The overall picture that emerges from our analysis is that in a compactification of type IIB string theory on an orientifold of a Calabi-Yau threefold hypersurface $X$ with $h^{1,1} \gg 1$, in the regime of control of the $\alpha'$ expansion, $X$ and most of its subvarieties have very large volumes in string units.  The resulting effective theory has many axions, some of which are essentially massless,\footnote{Many authors use the term `ultralight axion' for axions with $m \gtrsim 10^{-33}\,\eV$ that could make up part of the dark matter, as in \cite{Hu:2000ke,Hui:2016ltb}.
We avoid the term `ultralight' when speaking of the far lighter axions found here, with $m \ll 10^{-33}\,\eV$; these we instead call `massless', even though strictly speaking their masses are negligibly small, not zero.} with $m \ll 10^{-33}\,\eV$.  The axion kinetic matrix has small eigenvalues, and the radius of the axion fundamental domain is sub-Planckian.\footnote{As explained in \S\ref{sec:results}, in a small fraction of cases we cannot exclude the possibility of super-Planckian radii, but neither can we prove that all curves in $X$ have positive volume in these cases.  For the present discussion we consider only the better-established examples with $J \in \widetilde{\cK}_V$, for which the radii are sub-Planckian.}
In summary, we find an axiverse with \emph{hundreds of axions}, some of them \emph{massless}, and all with \emph{small periodicities}.  In this section we will mention a few of the implications of these findings for the cosmology of compactifications with $h^{1,1} \gg 1$.

Axions and axion-like particles with appreciable couplings to the Standard Model are strongly constrained by a wealth of data from diverse channels, including terrestrial appearance experiments such as helioscopes and haloscopes; red giant evolution; supernovae; CMB spectral distortions; and X-ray production in galactic or cosmological magnetic fields.  See \cite{Marsh:2015xka} for a review.  To apply these constraints to the large-$h^{1,1}$ axiverse that we have described, it would be necessary to make specific assumptions about the realization of the Standard Model, and its couplings to the axion sector.  While very interesting, such an analysis is extremely model-dependent.

Cosmological effects of the gravitational couplings of axions present a more direct, but still somewhat model-dependent, set of constraints.  Sufficiently light axions, with $m \ll 10^{-33}\,\eV$, are indistinguishable from vacuum energy unless excitations of the axion field (i.e., particles) are produced as dark radiation, for example through the decay of an associated modulus.
The limits on dark radiation are rather stringent, cf.~\cite{Cicoli:2012aq, Higaki:2012ar,Acharya:2015zfk}, but again depend on the details of post-inflationary evolution.  For example, if the lightest modulus decays only to a single axion, as well as to Standard Model particles, the dark radiation constraints are insensitive to the existence of other axions and moduli \cite{Acharya:2015zfk}, but can be severe nonetheless \cite{Cicoli:2012aq, Higaki:2012ar}.
Axions with $m \sim 10^{-33}\,\eV$ can be quintessence fields \cite{Frieman:1995pm}, and in special cases could even alleviate the ``why now'' problem \cite{Kamionkowski:2014zda}.
Axions with $m \sim 10^{-22}\,\eV$ could constitute a portion of the dark matter, and could give rise to small-scale structure in better agreement with observations than that predicted by cold dark matter models \cite{Hu:2000ke} (for recent work, see e.g.~\cite{Hui:2016ltb,Marsh:2013ywa,Schive:2014dra,Halverson:2018olu}).  Overproduction of axion dark matter --- and in some mass ranges, isocurvature perturbations in the CMB --- provide serious constraints \cite{Amendola:2005ad,Hlozek:2014lca,Hlozek:2017zzf}, especially in models with many axions \cite{Mack:2009hs}.

Perhaps the most interesting constraints on the large-$h^{1,1}$ axiverse come from black hole superradiance \cite{Arvanitaki:2009fg}.  Axions in the mass range $10^{-10}\,\eV-10^{-20}\,\eV$, even if not present as a cosmologically abundant population, can trigger instabilities of black holes.
Detailed modeling of moduli stabilization would be necessary to make precise statements, but as a rough estimate, we find that approximately half the geometries in our ensemble have an axion in the mass range $10^{-10}\,\eV \le m \le 10^{-20}\,\eV$.
Superradiance limits on many-axion theories have been obtained in \cite{Stott:2018opm}.
However, the analysis of \cite{Stott:2018opm} is only directly applicable to theories with relatively large decay constants, $f \gtrsim 10^{14}\,\mathrm{GeV}$.
Axions with smaller periodicities suffer from nonlinear interactions, potentially changing the limits of \cite{Stott:2018opm}.
A dedicated study of superradiance constraints on the Kreuzer-Skarke axiverse would be a worthwhile topic for the future.

\section{Conclusions}  \label{sec:conclusions}

We have initiated a survey
of compactifications on Calabi-Yau threefold hypersurfaces with arbitrary $h^{1,1}$, i.e.~of the entire Kreuzer-Skarke list.

This work extends and complements the complete enumeration carried out by Altman et al.~\cite{Altman} for $h^{1,1} \le 6$.  The large $h^{1,1}$ regime presents evident computational challenges, a few of which we have overcome.  Publicly-available software such as {\tt{Sage}} \cite{sagemath} generally produces FRSTs only for $h^{1,1} \lesssim 10$, and the improved triangulation algorithms that have been implemented on a large scale in the past are expensive, and function only for $h^{1,1} \lesssim 30$ \cite{HeavyTails,Altman}.  Moreover, the sheaf cohomology computations needed for studying divisors $D \subset X$ likewise explode in difficulty for $h^{1,1} \gtrsim 10$.  These limitations have led to a perception that systematic enumeration and study of hypersurfaces with $h^{1,1} \gg 10$ --- corresponding to the bulk of the Kreuzer-Skarke database --- is not possible at present.
In this work we have demonstrated, on the contrary, that large-scale studies are feasible for any range of $h^{1,1}$ arising in the Kreuzer-Skarke list, given only modest computational resources.

A key step was implementing the triangulation algorithm described in \cite{Braun:2014xka}, which allowed us to obtain fine regular \emph{star} triangulations $\sim 5\cdot 10^3$ times faster (per CPU) than was possible in \cite{HeavyTails,Altman}.  With our methods, finding one FRST takes just seconds even for $h^{1,1}=491$.\footnote{Given such a triangulation, the results of \cite{Puff} allow immediate study of the Hodge numbers of square-free divisors of the corresponding threefold.}
However, although we can efficiently generate large numbers of compactifications at any desired $h^{1,1}$, several challenges remain.
In this work, we used {\tt{Sage}} to obtain the intersection numbers of hypersurfaces with $h^{1,1} \le 100$, at a computational cost of order half a CPU-hour per hypersurface at $h^{1,1}=100$.  Significant further gains are possible in this area, and allow efficient computation of intersection numbers for any $h^{1,1} \le 491$, as we will show in \cite{Picardwipmisc}.
Even so, one thing that remains out of reach is a complete enumeration of hypersurfaces at large $h^{1,1}$, simply because the number of such hypersurfaces --- corresponding to the number of inequivalent triangulations of reflexive polytopes with many lattice points --- appears to be vast.

The principal raw data produced by our analysis are FRSTs
of four-dimensional reflexive polytopes; the K\"ahler cones of the corresponding toric varieties $V$; and the intersection numbers of generic Calabi-Yau threefold hypersurfaces $X \subset V$.  These data provide a wealth of information about four-dimensional effective theories arising from string compactifications on such $X$.
In this paper we studied two of the most salient physical findings, axion mass hierarchies and axion field ranges, leaving a complete characterization of the physical implications of our topological and geometric data as a significant task for future work.

The first
observable we studied is the set of relations among cycle volumes enforced by the Mori cone conditions, which control
the structure of the potential generated by instantons.  We found that enforcing that every effective curve has volume at least one in string units, as a proxy for ensuring control of the $\alpha'$ expansion, has --- for $h^{1,1} \gg 1$ --- a striking consequence: the volumes of many irreducible curves and
divisors on $X$, and of $X$ itself, become extremely large.
We found that these volumes scale roughly as $(h^{1,1})^p$, with the exponent $3 \lesssim p \lesssim 7$ depending on the type of cycle considered.

One consequence is that in a compactification of type IIB string theory on an orientifold of a typical Calabi-Yau threefold hypersurface with $h^{1,1} \gg 1$, one of the following holds:
\begin{enumerate}[1.]
\item{} One or more axions are effectively massless.
\item{} Many effective curves have volumes $\lesssim 1$.
\item{} The axion mass terms produced by Euclidean D3-branes wrapping non-holomorphic four-cycles are parametrically larger than those from holomorphic four-cycles.
\end{enumerate}
When condition (3) holds, the breakdown of the $\alpha'$ expansion cannot be detected by computing the volumes of calibrated cycles, while condition (2) suggests but does not guarantee the existence of large perturbative and nonperturbative corrections in the $\alpha'$ expansion.  Thus, we have established a tension between nonvanishing masses for all axions, and manifest control of the $\alpha'$ expansion.

The second observable we studied is the metric on K\"ahler moduli space, which is relevant for understanding quantum gravity constraints on large-field inflation.  We found that the eigenvalues of the axion kinetic matrix are typically small at large $h^{1,1}$, primarily because of the large volume of $X$. In each geometry we computed an approximation
to the radius
of the axion fundamental domain.  The radius depends strongly on how restrictive a condition one imposes on the K\"ahler form $J$.
For K\"ahler forms such that every curve in the ambient toric variety $V$ has volume $\ge 1$, we found field ranges
$\ll M_{\rm{pl}}$ in every example.  In the less restrictive case of K\"ahler forms in the region defined by \eqref{defkcap}, corresponding to the outer approximant to the stretched K\"ahler cone of $X$, we found super-Planckian axion field ranges in a small fraction of  geometries, at each $h^{1,1}$. While intriguing, this finding cannot be taken as evidence for large
field ranges in the regime of control of the $\alpha'$ expansion, because without a direct computation of $\cK_X$ we cannot exclude the possibility that in each example giving an apparent large field range, one or more effective curves $C \in \mathcal{M}_X$ has volume $<1$, or indeed $<0$.\footnote{Note that because $\cK_X \subset \cK_\cap$, computing $\cK_\cap$ is sufficient to place definite lower bounds on volumes, or upper bounds on field ranges.  However, because $\cK_X \subsetneq \cK_\cap$ in general, any examples with $J \not\in \cK_V$ are necessarily provisional, and await a direct computation of $\cK_X$.}  Overcoming this limitation is an important task for the future.

Because our results
are drawn from a statistical study of an ensemble of geometries, they should be taken as statements
about typical compactifications, and there can be special Calabi-Yau threefolds that violate the behavior observed in this work. For instance, the Mori
cone constraints in the geometry studied in~\cite{Denef:2005mm} are much milder than those of a typical hypersurface, possibly because the K\"ahler moduli space in~\cite{Denef:2005mm} has a symmetry corresponding to exchanging the K\"ahler moduli that correspond to blowups. It would be interesting to understand the incidence of such geometries.

Our results give a sharper picture of the spectrum of axion masses and decay constants arising in geometric compactifications of string theory.  Reasonable a priori estimates of these spectra, as well as studies in families of examples, have been made and used in the study of the string axiverse \cite{Arvanitaki:2009fg,Acharya:2010zx,Cicoli:2012sz,Acharya:2015zfk,Stott:2017hvl}, and our ensemble provides a foundation for refining these estimates.

There are several
directions for future work.  By applying computational resources on a larger scale, one could more finely sample the Kreuzer-Skarke database.  A rough estimate is that in under a few hundred CPU-years one could find \emph{one} FRST for every one of the 473,800,776 polytopes in the list.  Intersection numbers could be obtained at comparable cost given the improved methods of \cite{Picardwipmisc}.  As we have stressed, however, the number of distinct triangulations is plausibly vastly larger than the number of polytopes, and so it remains to be seen whether it is possible even
to store the topological
data of every compactification arising from the Kreuzer-Skarke database.

The geometric data obtained here can serve to answer questions about which sorts of effective theories are possible in compactifications on Calabi-Yau hypersurfaces.
To answer such questions, it would be natural to use machine learning~\cite{Abel:2014xta, He:2017aed, Ruehle:2017mzq, Krefl:2017yox, Carifio:2017bov, Carifio:2017nyb, Wang:2018rkk, He:2018dlv}, among other tools, given the scale and complexity of the data.

\section*{Acknowledgments}

We thank Ross Altman, Thomas Bachlechner, Mike Douglas, Thomas Grimm, Jim Halverson, Arthur Hebecker, Lionel Levine, Eran Palti, John Stout, and Alexander Westphal for discussions.  We are indebted to Doddy Marsh for a careful reading of a draft, and for guidance on the material in \S\ref{sec:implications}.  We thank Benjamin Sung for sharing code for computing the K\"ahler cone of a toric variety.
The work of M.D.~and L.M.~was supported in part by NSF grant PHY-1719877.  The work of C.L.~was supported by NSF grant PHY-1620526.
The work of M.S.~was supported in part by NSF grant DMS-1502294.

\appendix

\newpage

\section{Non-holomorphic Instantons}  \label{sec:nonholomorphic}

We noted in \S\ref{sec:background} that computing contributions to the superpotential from Euclidean D3-branes wrapping holomorphic four-cycles (i.e., effective divisors)
is much simpler than computing contributions to the K\"ahler potential from Euclidean D3-branes wrapping non-holomorphic four-cycles.  On the physics side, one reason for the disparity is that superpotential terms are constrained by holomorphy.  Geometrically, the difference between the two computations is that a holomorphic four-cycle $\Sigma_4$ is calibrated by the K\"ahler form $J$, and enjoys the relation
\begin{equation}
\mathrm{Vol}(\Sigma_4) = \frac{1}{2}\int_{\Sigma_4}J\wedge J\,,
\end{equation} so that once $J$ is given, $\mathrm{Vol}(\Sigma_4)$ is determined by topological data.  Similarly, an \emph{antiholomorphic} four-cycle $\overline{\Sigma}_4$ has orientation opposite to that of a holomorphic cycle, and
obeys
\begin{equation}
\mathrm{Vol}(\overline{\Sigma}_4) = -\frac{1}{2}\int_{\overline{\Sigma}_4}J\wedge J\,,
\end{equation}
However, it is much more difficult to compute the volume of a cycle that is neither holomorphic nor antiholomorphic, as we now explain.

\subsection{Volume-minimizing currents}

Suppose that $X$ is a compact K\"ahler manifold of dimension $n$,\footnote{Assuming that $X$ is Calabi-Yau, and/or that $n=3$, does not lead to appreciable simplifications, and so we shall not make these assumptions in this section.} with K\"ahler form $J$, and consider a class $[\alpha] \in H_{2n-2}(X,\mathbb{Z})$.
By definition, $[\alpha]$ can be represented by some effective divisor $D$ if and only if
$[\alpha] \in \mathrm{Eff}(X)$.  So suppose, henceforth, that $[\alpha] \not\in \mathrm{Eff}(X)$, and also $-[\alpha] \not\in \mathrm{Eff}(X)$.  Then $[\alpha]$ admits neither a holomorphic representative nor an antiholomorphic representative.

Writing $\mathrm{Vol}(\alpha)$ for the volume of a given representative $\alpha$ of the class $[\alpha]$, one might attempt to define
\begin{equation}\label{minvoldef}
\mathrm{MinVol}([\alpha])= \displaystyle \min_{\alpha \in [\alpha]} \mathrm{Vol}(\alpha)\,,
\end{equation}
i.e.~$\mathrm{MinVol}([\alpha])$ is the volume of the smallest-volume representative of the class $\alpha$.

However, it is not obvious that the variational problem implied by \eqref{minvoldef}
is well-posed: does one search over all representatives of $\alpha$, or just representatives obeying an appropriate smoothness condition?  It is also not clear a priori how smooth the volume-minimizing configuration will be: in fact, one can easily find examples in which the volume-minimizing configuration has singularities, at least at complex codimension one.

Fortunately, the problem of finding the minimum-volume representative of a given homology class is one of the central questions of \emph{geometric measure theory}, and was put on sound footing in the 1960s by Federer and Fleming.  They defined objects called integral $p$-currents, which roughly correspond to formal sums of $p$-dimensional submanifolds, except for sets of $p$-dimensional Hausdorff measure zero.
Federer and Fleming showed that the class of integral $p$-currents has a compactness property that is very useful in formulating variational problems: in fact, they proved that for each class\footnote{Federer and Fleming's theory of integral currents is is not limited to the case that $X$ is K\"ahler, nor even complex, nor does it require that $[\alpha]$ is dual to a hypersurface, but for simplicity of presentation we state what their results imply for the case of present interest.} $[\alpha] \in H_{2n-2}(X,\mathbb{Z})$, there exists an integral current of least volume \cite{FF}.  In other words, \eqref{minvoldef} actually does define the solution of a well-posed variational problem, provided that $\alpha$ is understood to vary over integral currents, not just over smooth submanifolds.

\subsection{Non-holomorphic instantons and volume reduction}

Consider a Euclidean D3-brane in a homology class $[\alpha] \not\in \mathrm{Eff}(X)$, which necessarily
cannot contribute to the superpotential, but may contribute to the K\"ahler potential.
The real part of the action of such a Euclidean D3-brane is plausibly proportional to $\mathrm{MinVol}([\alpha])$, which is well-defined thanks to geometric measure theory.  Even so, computing $\mathrm{MinVol}([\alpha])$ is nontrivial.

As a toy example, suppose that $X$ is such that four-cycles $\alpha_1$ and $\alpha_2$ are a basis for $H_{4}(X,\mathbb{Z})=\mathbb{Z}^2$, and $\alpha_1$ and $\alpha_2$ also generate $\mathrm{Eff}(X)$.  For a given K\"ahler form $J$, define $\tau_i:=\frac{1}{2}\int_{\alpha_i}J\wedge J$ and $\theta_i:=\int_{\alpha_i}C_4$, $i=1,2$.  If $J$ is such that $\tau_1,\tau_2 \gg L$ for some $L \gg 1$, then Euclidean D3-brane terms in the superpotential are no larger than $e^{-2\pi L}$.

In this situation, one should ask about contributions to the K\"ahler potential from Euclidean D3-branes wrapping a representative $\gamma$ of a non-effective class such as $[\gamma]:=[\alpha_1-\alpha_2] \not\in \mathrm{Eff}(X)$.  Because $\alpha_1$ and $\alpha_2$ are calibrated by $J$, we have $\mathrm{MinVol}([\alpha_i])=\tau_i \gg L$.
The action of instantons on $\gamma$ is determined by $\tau_{\gamma}:=\mathrm{MinVol}([\gamma])$.
However, we cannot conclude that
\begin{equation} \label{dotheyadd}
\tau_{\gamma} \ge \tau_{1} + \tau_{2}\,.
\end{equation}
If $\alpha_1$ and $\alpha_2$ are disjoint, then \eqref{dotheyadd} actually does hold, but more generally the intersection locus of the minimum-volume representatives of $[\alpha_1]$ and $[\alpha_2]$ can be deformed to produce a representative of $[\gamma]$ with volume $<\tau_{1} + \tau_{2}$.
When $\tau_{\gamma} = \tau_{1} + \tau_{2} - \Delta\tau$ for $\Delta\tau>0$, we will say that recombination has led to \emph{volume reduction} by an amount $\Delta\tau>0$.

The question of volume reduction is best-understood for two-dimensional currents.  Building on work of Almgren \cite{Almgren}, Chang proved that in any Riemannian manifold, the singular set of a volume-minimizing two-dimensional current consists of isolated branch points \cite{Chang}.  It is therefore tempting to conjecture that in a K\"ahler manifold, a volume-minimizing two-dimensional current consists of a union of holomorphic and antiholomorphic curves, intersecting at points; one consequence would be that there is no volume reduction for
two-dimensional currents in a K\"ahler manifold.  However, in \cite{Micallef}, for $X$ a K3 surface, Micallef and Wolfson gave an explicit example of a non-effective class $[\alpha_1-\alpha_2] \in H_{2}(K3,\mathbb{Z})$ whose minimum-volume representative is \emph{not} a union of holomorphic and antiholomorphic curves,\footnote{However, see \cite{Arezzo} for a related variational problem whose extrema are unions of holomorphic and antiholomorphic curves.} and for which
$\mathrm{MinVol}([\alpha_1-\alpha_2])<\mathrm{MinVol}([\alpha_1])+\mathrm{MinVol}([\alpha_2])$.
The volume reduction is proportional to the small parameter $\varepsilon$ measuring the deviation from the orbifold limit of K3.

The issue, returning to four-cycles, is then the following.  If a significant volume reduction $\Delta\tau \sim \mathcal{O}(\tau_1,\tau_2)$ could occur in some setting, so that $\tau_{\gamma} \ll \tau_i$, then ensuring $\tau_1,\tau_2 \gg L$ would not place any upper bound on the size of Euclidean D3-brane terms in the K\"ahler potential.
The axion masses from non-holomorphic instantons in $K$ would be parametrically larger than those from holomorphic instantons in $W$.

Although the Micallef-Wolfson construction proves that nonzero volume reduction can occur in a Calabi-Yau compactification, we are not aware of any example of parametrically large volume reduction in a comparable setting.  Moreover, the cycle volume determines only the leading semiclassical action of a Euclidean D-brane, and one should compute corrections to this action, such as the fluctuation determinant, before drawing conclusions about the relative sizes of physical effects.\footnote{We thank Eran Palti for comments on this point.}

In summary, determining whether Euclidean D3-branes wrapping non-holomorphic cycles can contribute axion masses that are parametrically larger than those arising from holomorphic cycles is an open problem.  The available evidence does not exclude this possibility, but also does not, in our view, strongly support it.  Our results on axion masses rely on our computation of the volumes of holomorphic cycles, and could be affected if large volume reduction occurs and causes non-holomorphic instantons to dominate in the potential.  This proviso should be kept in mind when interpreting our findings.

\FloatBarrier

\bibliographystyle{JHEP}
\bibliography{refs}

\end{document}